\documentclass[a4paper,11pt]{article}
\pdfoutput=1 
\usepackage{jcappub} 
\usepackage{lmodern} 
\usepackage[T1]{fontenc}
\usepackage{amsmath}
\usepackage{amssymb}
\def\h2o{\ifmmode x_{\rm H_2O} \else $x_{\rm H_2O}$ \fi}
\def\iuv{\ifmmode I_{\rm UV} \else $I_{\rm UV}$ \fi}
\def\in{\ifmmode I_{\rm UV}/n_4 \else $I_{\rm UV}/n_4$ \fi}
\def\mw{\ifmmode x_{\rm H_2O, MW} \else $x_{\rm H_2O, MW}$ \fi}
\newcommand\hh{\ifmmode {\rm H_2} \else H$_2$ \fi}
\def\no{\ifmmode {N_{\rm HI}} \else $N_{\rm HI}$ \fi}
\def\nt{\ifmmode {N_{\rm H_2}} \else $N_{\rm HI}$ \fi}
\def\so{\ifmmode {\Sigma_{\rm HI}} \else $\Sigma_{\rm HI}$ \fi}
\def\st{\ifmmode {\Sigma_{\rm H_2}} \else $\Sigma_{\rm H_2}$ \fi}
\renewcommand\ss{\ifmmode {\Sigma_{\rm tot}} \else $\Sigma_{\rm tot}$ \fi}
\def\msun{\ifmmode {\rm M_{\odot}}\else $\rm M_{\odot}$\fi}
\def\mpc{\ifmmode {\rm M_{\odot}/pc^2}\else $\rm M_{\odot}/pc^2$\fi}
\def\tra{\ifmmode {\rm HI-to-H_2}\else H{\small I}-to-H$_2$ \fi}

\title{On the Habitability of Our Universe}

\author{Abraham Loeb}

\affiliation{Astronomy department, Harvard University, 60 Garden
  Street, Cambridge, MA 02138, USA} 

\emailAdd{aloeb@cfa.harvard.edu}

\abstract{Is life most likely to emerge at the present cosmic time
  near a star like the Sun? We consider the habitability of the
  Universe throughout cosmic history, and conservatively restrict our
  attention to the context of ``life as we know it'' and the standard
  cosmological model, $\Lambda$CDM.  The habitable cosmic epoch
  started shortly after the first stars formed, about 30 Myr after the
  Big Bang, and will end about 10 Tyr from now, when all stars will
  die. We review the formation history of habitable planets and find
  that unless habitability around low mass stars is suppressed, life
  is most likely to exist near $\sim 0.1M_\odot$ stars ten trillion
  years from now. Spectroscopic searches for biosignatures in the
  atmospheres of transiting Earth-mass planets around low mass stars
  will determine whether present-day life is indeed premature or
  typical from a cosmic perspective.}

\begin{document}

\maketitle

\section{Introduction}
\label{Sec:intro}

The known forms of terrestrial life involve carbon-based chemistry in
liquid water~\cite{Kasting1,Kasting}. In the cosmological context, life could
not have started earlier than 10 Myr after the Big Bang ($z\gtrsim
140$) since the entire Universe was bathed in a thermal radiation
background above the boiling temperature of liquid water. Later on,
however, the Universe cooled to a {\it habitable epoch} at a
comfortable temperature of 273-373 K between 10-17 Myr after the Big
Bang~\cite{Loeb14}.

The phase diagram of water allows it to be liquid only under external
pressure in an atmosphere which can be confined gravitationally on the
surface of a planet. To keep the atmosphere bound against evaporation
requires strong surface gravity of a rocky planet with a mass
comparable to or above that of the Earth~\cite{SchallerBrown}.

The emergence of ``life as we know it'' requires stars for two
reasons. Stars are needed to produce the heavy elements (carbon,
oxygen and so on, up to iron) out of which rocky planets and the
molecules of life are made. Stars also provide a heat source for
powering the chemistry of life on the surface of their planets. Each
star is surrounded by a habitable zone where the surface temperature
of a planet allows liquid water to exist. The approximate distance of
the habitable zone, $r_{\rm HZ}$, is obtained by equating the heating
rate per unit area from the stellar luminosity, $L$, to the cooling
rate per unit area at a surface temperature of $T_{\rm HZ}\sim 300$~K,
namely $({L/ 4\pi r_{\rm HZ}^2})\sim \sigma T_{\rm HZ}^4$, where
$\sigma$ is the Stefan-Boltzman constant~\cite{Kasting, Kasting1}.
%The habitable zone is commonly defined in reference to a distance from
%a luminous source, such as a star~\cite{Kasting,Kasting1}, whose heat
%maintains the surface of a rocky planet at a temperature of $\sim
%300$K, allowing liquid water to exist and the chemistry of ``life as
%we know it'' to operate.  

Starting from the vicinity of particular stars, life could potentially
spread. This process, so-called panspermia, could be mediated by the
transfer of rocks between planetary systems~\cite{SpAd}.  The
``astronauts'' on such rocks could be microscopic animals such as {\it
  tardigrades}, which are known to be resilient to extreme vacuum,
dehydration and exposure to radiation that characterize space
travel. In 2007 dehydrated tardigrades were taken into a low Earth
orbit for ten days, and after returning to Earth - most of them
revived after rehydration and many produced viable
embryos.~\footnote{http://www.bbc.com/earth/story/20150313-the-toughest-animals-on-earth}
Life which arose via spreading will exhibit more clustering than life
which arose spontaneously, and so the existence of panspermia can be
detected statistically through excess spatial correlations of life
bearing environments. Future searches for biosignatures in the
atmospheres of exoplanets could test for panspermia: a smoking gun
signature would be the detection of large regions in the Milky Way
where life saturates its environment interspersed with voids where
life is very uncommon.  In principle, detection of as few as several
tens of biologically active exoplanets could yield a highly
significant detection of panspermia~\cite{LinLoeb}. Once life emerges
on the surface of a planet, it is difficult to extinguish it
completely through astrophysical events (such as quasar activity,
supernovae, gamma-ray bursts, or asteroid impacts) other than the
death of the host star. Life is known to be resilient to survival in
extreme environments and could be protected from harmful radiation or
heat if it resides underground or in the deep ocean floor.

Panspermia is not limited to galactic scales and could extend over
cosmological distances. Some stars and their planets are ejected from
their birth galaxies at a speed approaching the speed of light,
through gravitational slingshot from pairs of supermassive black holes
which are formed during galaxy mergers~\cite{LG15,GL15}. The resulting
population of relativistic stars roaming through intergalactic space
could potential transfer life between galaxies separated by vast
distances across the Universe.

The spread of life could be enhanced artificially through the use of
spacecrafts by advanced civilizations. Our own civilization is
currently starting to develop the technology needed to visit the
nearest stars with a travel time of decades through the propulsion of
lightweight sails to a fraction of the speed of light by a powerful
laser.\footnote{http://www.breakthroughinitiatives.org/Concept/3} The
existence of advanced civilizations could be revealed through the
detection of industrial pollution in the atmospheres of
planets~\cite{LGL}, the detection of powerful beacons of light used
for propulsion~\cite{SEGL} and communications~\cite{ZL}, or through
artificial lights~\cite{TL12}. The search for signatures of advanced
civilizations is the richest interdisciplinary frontier, offering
interfaces between astronomy and other disciplines, such as biology
(astro-biology), chemistry (astro-chemistry), statistics
(astro-statistics), or engineering (astro-engineering). Moreover, the
prospects for communication with aliens could open new disciplines on
the interface with linguistics (astro-linguistics), psychology
(astro-psychology), sociology (astro-sociology), philosophy
(astro-philosophy) and many other fields. {\it ``Are we alone?''} is
one of the most fundamental questions in science; the answer we find
to this question is likely to provide a fresh perspective on our place
in the Universe. Although primitive forms of life are likely to be
more abundant, intelligent civilizations could make themselves
detectable out to greater distances.

The search for life was reinvigorated by the Kepler satellite which
revealed that a substantial fraction of the stars in the Milky Way
galaxy host habitable Earth-mass planets around
them~\cite{EtaEarthSilburt,Petigura,Marcy,EtaEarthDC}. Indeed, the
nearest star to the Sun, Proxima, whose mass is only 12\% of the solar
mass, was recently found to host an Earth-mass planet in the habitable
zone. This planet, Proxima b, is twenty times closer to its faint
stellar host than the Earth is to the Sun~\cite{Ang}. Dwarf stars like
Proxima are the most abundant stars in the Universe and they live for
trillions of years, up to a thousand times longer than the Sun. If
life could form around them, it would survive long into the
future. The prospects for life in the distant cosmic future can
therefore be explored by searching for biosignatures around nearby
dwarf stars. For example, the existence of an atmosphere around
Proxima's planet could be detected relatively soon by measuring the
temperature contrast between its day and night sides~\cite{KL16}.
Speaking about the cosmic future, it is interesting to note that
planets or rocky debris are known to exist around stellar remnants,
such as white dwarfs~\cite{Coper,LL15} and neutron
stars~\cite{Ray16}. The Sun is currently at the middle of its lifetime
on its way to become a white dwarf.  White dwarfs which are billions
of years old and exist in abundance comparable to that of Sun-like
stars, have a surface temperature similar to that of the Sun but are a
hundred times smaller in size. As a result, the habitable zone around
them is a hundred times closer than the Earth is to the
Sun~\cite{Agol}. Searches for biosignatures in the atmospheres of
habitable planets which transit white dwarfs could potentially be
conducted in the near future~\cite{LoebMaoz}. Counting all possible
host stars and extrapolating to cosmological scales, there might be at
present as many as $\sim 10^{20}$ habitable planets in the observable
volume of the Universe~\cite{Beh,Zac}.

In the following sections of this chapter we discuss the habitability
of the Universe as a function of cosmic time, starting from the
earliest habitable epoch in \S \ref{habep}.  According to the standard
model of cosmology, the first stars in the observable Universe formed
$\sim 30$ Myr after the Big Bang at a redshift, $z\sim
70$~\cite{Loeb14,LF13,Fialkov,Naoz}. Within a few Myr, the first
supernovae dispersed heavy elements into the surrounding gas,
enriching the second generation stars with heavy elements. Remnants
from the second generation of stars are found in the halo of the Milky
Way galaxy, and may have planetary systems in the habitable zone
around them~\cite{MashianLoeb}, as discussed in \S \ref{cemphalo}. The
related planets are likely made of carbon, and water could have been
delivered to their surface by icy comets, in a similar manner to the
solar system. The formation of water is expected to consume most of
the oxygen in the metal poor interstellar medium of the first
galaxies~\cite{Bialy}, as discussed in \S \ref{waterfirst}.
Therefore, even if the cosmological constant was bigger than its
measured value by up to a factor of $\sim 10^3$ so that galaxy
formation was suppressed at redshifts $z\lesssim 10$, life could have
still emerged in our Universe due to the earliest generation of
galaxies~\cite{Loeb06}, as discussed in \S \ref{anth}.  We conclude in
\S \ref{likelihood} with a calculation of the relative likelihood per
unit time for the emergence of life~\cite{LBS}, which is of particular
importance for studies attempting to gauge the level of fine-tuning
required for the cosmological or fundamental physics parameters that
would allow life to emerge in our Universe.

\section{The Habitable Epoch of the Early Universe}
\label{habep}

\subsection{Section Background}

We start by pointing out that the cosmic microwave background (CMB)
provided a uniform heating source at a temperature of $T_{\rm
  cmb}=272.6 {\rm K} \times [(1+z)/100]$ \cite{Fixsen} that could have
made by itself rocky planets habitable at redshifts $(1+z)=100$--137
in the early Universe, merely 10--17 million years after the Big Bang.

In order for rocky planets to exist at these early times, massive
stars with tens to hundreds of solar masses, whose lifetime is much
shorter than the age of the Universe, had to form and enrich the
primordial gas with heavy elements through winds and supernova
explosions~\cite{Ober,Heger}. Indeed, numerical simulations predict
that predominantly massive stars have formed in the first halos of
dark matter to collapse~\cite{BL04,LF13}.  For massive stars that are
dominated by radiation pressure and shine near their Eddington
luminosity $L_{\rm E}=1.3\times 10^{40}~{\rm
  erg~s^{-1}}(m/100M_\odot)$, the lifetime is independent of stellar
mass $m$ and set by the 0.7\% nuclear efficiency for converting rest
mass to radiation, $\sim (0.007m c^2)/L_{\rm E}= 3~{\rm
  Myr}$~\cite{El,BKL}. We next examine how early did such stars form
within the observable volume of our Universe.

\subsection{First Planets}

In the standard cosmological model, structure forms hierarchically --
starting from small spatial scales, through the gravitational growth
of primordial density perturbations~\cite{LF13}. On any given spatial
scale $R$, the probability distribution of fractional density
fluctuations $\delta$ is assumed to have a Gaussian form,
$P(\delta)d\delta
=(2\pi\sigma^2)^{-1/2}\exp\{-\delta^2/2\sigma^2\}d\delta$, with a
{\it root-mean-square} amplitude $\sigma(R)$ that is initially much
smaller than unity.  The initial $\sigma(R)$ is tightly constrained on
large scales, $R\gtrsim 1~{\rm Mpc}$, through observations of the CMB
anisotropies and galaxy surveys~\cite{Planck,And}, and is
extrapolated theoretically to smaller scales. Throughout the discussion, we
normalize spatial scales to their so-called ``comoving'' values in the
present-day Universe. The assumed Gaussian shape of $P(\delta)$ has so
far been tested only on scales $R\gtrsim 1~{\rm Mpc}$ for $\delta
\lesssim 3\sigma$~\cite{Shan}, but was not verified in the far tail
of the distribution or on small scales that are first to collapse in
the early Universe.

As the density in a given region rises above the background level, the
matter in it detaches from the Hubble expansion and eventually
collapses owing to its self-gravity to make a gravitationally bound
(virialized) object like a galaxy. The abundance of regions that
collapse and reach virial equilibrium at any given time depends
sensitively on both $P(\delta)$ and $\sigma(R)$. Each collapsing
region includes a mix of dark matter and ordinary matter (often
labeled as ``baryonic'').  If the baryonic gas is able to cool below
the virial temperature inside the dark matter halo, then it could
fragment into dense clumps and make stars.

At redshifts $z\gtrsim 140$ Compton cooling on the CMB is effective on
a timescale comparable to the age of the Universe, given the residual
fraction of free electrons left over from cosmological recombination
(see \S 2.2 in Ref.~\cite{LF13} and also Ref.~\cite{PL12}). The
thermal coupling to the CMB tends to bring the gas temperature to
$T_{\rm cmb}$, which at $z\sim 140$ is similar to the temperature
floor associated with molecular hydrogen
cooling~\cite{Haiman,Tegmark,Hirata}. In order for virialized gas in a
dark matter halo to cool, condense and fragment into stars, the halo
virial temperature $T_{\rm vir}$ has to exceed $T_{\rm min}\approx
300$K, corresponding to $T_{\rm cmb}$ at $(1+z)\sim 110$. This implies
a halo mass in excess of $M_{\rm min}=10^4 M_\odot$, corresponding to
a baryonic mass $M_{\rm b, min}=1.5\times 10^3~M_\odot$, a circular
virial velocity $V_{\rm c, min}=2.6~{\rm km~s^{-1}}$ and a virial
radius $r_{\rm vir,min }=6.3~{\rm pc}$ (see \S 3.3 in
Ref.~\cite{LF13}). This value of $M_{\rm min}$ is close to the minimum
halo mass to assemble baryons at that redshift (see \S 3.2.1 in
Ref.~\cite{LF13} and Fig. 2 of Ref.~\cite{TBH}).

The corresponding number of star-forming halos on our past light cone
is given by~\cite{Naoz},
\begin{equation}
N=\int_{(1+z)=100}^{(1+z)=137} n(M>M_{\rm min},z^\prime) {dV\over
dz^\prime} dz^\prime ,
\end{equation}
where $n(M>M_{\rm min})$ is the comoving number density of halos with
a mass $M>M_{\rm min}$~\cite{Sheth}, and $dV=4\pi r^2dr$ is the
comoving volume element with $dr=cdt/a(t)$. Here, $a(t)=(1+z)^{-1}$ is
the cosmological scale factor at time $t$, and $r(z)=c\int_0^z
dz^\prime/H(z^\prime)$ is the comoving distance. The Hubble parameter
for a flat Universe is $H(z)\equiv (\dot{a}/a)
=H_0\sqrt{\Omega_m(1+z)^3+\Omega_r(1+z)^4+\Omega_\Lambda}$, with
$\Omega_m$, $\Omega_r$ and $\Omega_\Lambda$ being the present-day
density parameters of matter, radiation and vacuum, respectively. The
total number of halos that existed at $(1+z)\sim 100$ within our
entire Hubble volume (not restricted to the light cone), $N_{\rm
tot}\equiv n (M>M_{\rm min},z=99)\times (4\pi/3)(3c/H_0)^3$, is larger
than $N$ by a factor of $\sim 10^3$.

For the standard cosmological parameters~\cite{Planck}, we find that
the first star-forming halos on our past light cone reached its
maximum turnaround radius\footnote{In the spherical collapse model,
the turnaround time is half the collapse time.} (with a density
contrast of 5.6) at $z\sim 112$ and collapsed (with an average density
contrast of 178) at $z\sim 71$. Within the entire Hubble volume, a
turnaround at $z\sim 122$ resulted in the first collapse at $z\sim
77$.  This result includes the delay by $\Delta z\sim 5.3$ expected
from the streaming motion of baryons relative to the dark matter
\cite{Fialkov}.

The above calculation implies that rocky planets could have formed
within our Hubble volume by $(1+z)\sim 78$ but not by $(1+z)\sim 110$
if the initial density perturbations were perfectly Gaussian. However,
the host halos of the first planets are extremely rare, representing
just $\sim 2\times 10^{-17}$ of the cosmic matter inventory. Since
they lie $\sim 8.5$ standard deviations ($\sigma$) away on the
exponential tail of the Gaussian probability distribution of initial
density perturbations, $P(\delta)$, their abundance could have been
significantly enhanced by primordial non-Gaussianity
\cite{LoVerde,Maio,Musso} if the decline of $P(\delta)$ at high values of
$\delta/\sigma$ is shallower than exponential. The needed level of
deviation from Gaussianity is not ruled out by existing data sets
\cite{Planck2}. Non-Gaussianity below the current limits is expected
in generic models of cosmic inflation~\cite{Malda} that are commonly
used to explain the initial density perturbations in the Universe.

\subsection{Section Summary and Implications}

In the discussion thus far, we highlighted a new regime of
habitability made possible for $\sim 6.6$ Myr by the uniform CMB
radiation at redshifts $(1+z)= 100$--137, just when the first
generation of star-forming halos (with a virial mass $\gtrsim
10^4M_\odot$) turned around in the standard cosmological model with
Gaussian initial conditions. Deviations from Gaussianity in the far
($8.5\sigma$) tail of the probability distribution of initial density
perturbations, could have led already at these redshifts to the birth
of massive stars, whose heavy elements triggered the formation of
rocky planets with liquid water on their surface.\footnote{The
  dynamical time of galaxies is shorter than $\sim 1/\sqrt{200}= 7\%$
  of the age of the Universe at any redshift since their average
  density contrast is $\gtrsim 200$. After the first stars formed, the
  subsequent delay in producing heavy elements from the first
  supernovae could have been as short as a few Myr. The supernova
  ejecta could have produced high-metallicity islands that were not
  fully mixed with the surrounding primordial gas, leading to
  efficient formation of rocky planets within them.}

Thermal gradients are needed for life. These can be supplied by
geological variations on the surface of rocky planets. Examples for
sources of free energy are geothermal energy powered by the planet's
gravitational binding energy at formation and radioactive energy from
unstable elements produced by the earliest supernova. These internal
heat sources (in addition to possible heating by a nearby star), may
have kept planets warm even without the CMB, extending the habitable
epoch from $z\sim 100$ to later times.  The lower CMB temperature at
late times may have allowed ice to form on objects that delivered
water to a planet's surface, and helped to maintain the cold trap of
water in the planet's stratosphere.  Planets could have kept a blanket
of molecular hydrogen that maintained their warmth
\cite{Stevenson,Gaidos}, allowing life to persist on internally warmed
planets at late cosmic times. If life persisted at $z\lesssim 100$, it
could have been transported to newly formed objects through
panspermia. Under the assumption that interstellar panspermia is
plausible, the redshift of $z\sim 100$ can be regarded as the earliest
cosmic epoch after which life was possible in our Universe.

Finally, we note that an increase in the initial amplitude of density
perturbations on the mass scale of $10^4M_\odot$ by a modest factor of
$1.4\times [(1+z)/110]$ would have enabled star formation within the
Hubble volume at redshifts $(1+z)>110$ even for perfectly Gaussian
initial conditions.  

\section{CEMP Stars: Possible Hosts to Carbon Planets in the Early Universe}
\label{cemphalo}

\subsection{Section Background}

The questions of when, where, and how the first planetary systems
actually formed in cosmic history remain crucial to our understanding
of structure formation and the emergence of life in the early
Universe~\cite{Loeb14}.
%In the Cold Dark Matter model of hierarchical
%structure formation, the first stars are predicted to have formed in
%dark matter haloes that collapsed at redshifts $z \lesssim$ 70, about
%30 million years after the Big
%Bang~\cite{Tegmark1,2001PhR...349..125B,2003ApJ...592..645Y,BL04,LF13}.
The short-lived, metal-free, massive first-generation stars ultimately
exploded as supernovae (SNe) and enriched the interstellar medium
(ISM) with the heavy elements fused in their cores. The enrichment of
gas with metals that had otherwise been absent in the early Universe
enabled the formation of the first low-mass stars, and perhaps, marked
the point at which star systems could begin to form
planets~\cite{2003Natur.425..812B,2007MNRAS.380L..40F,2008ApJ...672..757C}. In
the core accretion model of planet formation
(e.g. ~\cite{2006RPPh...69..119P,2011ApJ...736...89J}), elements
heavier than hydrogen and helium are necessary not only to form the
dust grains that are the building blocks of planetary cores, but to
extend the lifetime of the protostellar disk long enough to allow the
dust grains to grow via merging and accretion to form planetesimals
\cite{2005AA...430.1133K,2009ApJ...704L..75J,2009ApJ...705...54Y,2010MNRAS.402.2735E}.

In the past four decades, a broad search has been launched for
low-mass Population II stars in the form of extremely metal-poor
sources within the halo of the Galaxy. The HK survey
\cite{1985AJ.....90.2089B}, the Hamburg/ESO
Survey~\cite{1996AAS..115..227W,2008AA...484..721C}, the Sloan
Digital Sky Survey (SDSS; ~\cite{2000AJ....120.1579Y}), and the
SEGUE survey~\cite{2009AJ....137.4377Y} have all significantly
enhanced the sample of metal-poor stars with [Fe/H] $<$
--2.0. Although these iron-poor stars are often referred to in the
literature as ``metal-poor" stars, it is critical to note that [Fe/H]
does not necessarily reflect a stellar atmosphere's total metal
content. The equivalence between ``metal-poor" and ``Fe-poor" appears
to fall away for stars with [Fe/H] $<$ --3.0 since many of these stars
exhibit large overabundances of elements such as C, N, and O; the
total mass fractions, $Z$, of the elements heavier than He are
therefore not much lower than the solar value in these iron-poor
stars.

Carbon-enhanced metal-poor (CEMP) stars comprise one such chemically
anomalous class of stars, with carbon-to-iron ratios [C/Fe] $\geq$ 0.7
(as defined in
~\cite{2007ApJ...655..492A,2012ApJ...744..195C,2013ApJ...762...28N}). The
fraction of sources that fall into this category increases from
$\sim$15-20\% for stars with [Fe/H] $<$ --2.0, to 30\% for [Fe/H] $<$
--3.0, to $\sim$75\% for [Fe/H] $<$ --4.0
\cite{2005ARAA..43..531B,2013ApJ...762...28N,2015ARAA..53..631F}. Furthermore,
the degree of carbon enhancement in CEMP stars has been shown to
notably increase as a function of decreasing metallicity, rising from
[C/Fe] $\sim$ 1.0 at [Fe/H] = -1.5 to [C/Fe] $\sim$ 1.7 at [Fe/H] =
-2.7.~\cite{2012ApJ...744..195C}. Given the significant frequency and
level of carbon-excess in this subset of metal-poor Population II
stars, the formation of carbon planets around CEMP stars in the early
universe presents itself as an intriguing possibility.

From a theoretical standpoint, the potential existence of carbon
exoplanets, consisting of carbides and graphite instead of Earth-like
silicates, has been suggested by Ref.~\cite{2005astro.ph..4214K}. Using
the various elemental abundances measured in planet-hosting stars,
subsequent works have sought to predict the corresponding variety of
terrestrial exoplanet compositions expected to exist
\cite{2010ApJ...715.1050B,2012ApJ...747L...2C,2012ApJ...760...44C}. Assuming
that the stellar abundances are similar to those of the original
circumstellar disk, related simulations yield planets with a whole
range of compositions, including some that are almost exclusively C
and SiC; these occur in disks with C/O $>$ 0.8, favorable conditions
for carbon condensation~\cite{1975GeCoA..39..389L}. Observationally,
there have also been indications of planets with carbon-rich
atmospheres, e.g. WASP-12b~\cite{2011Natur.469...64M}, and
carbon-rich interiors, e.g. 55 Cancri e~\cite{2012ApJ...759L..40M}.

In this section, we explore the possibility of carbon planet formation
around the iron-deficient, but carbon-rich subset of low-mass stars,
mainly, CEMP stars. Standard definitions of elemental abundances and
ratios are adopted. For element X, the logarithmic absolute abundance
is defined as the number of atoms of element X per 10$^{12}$ hydrogen
atoms,
$\log{\epsilon(\textrm{X})}=\log_{10}{(N_{\textrm{X}}/N_{\textrm{Y}})}+12.0.$
For elements X and Y, the logarithmic abundance ratio relative to the
solar ratio is defined as [X/Y] =
$\log_{10}{(N_{\textrm{X}}/N_{\textrm{Y}})}-\log_{10}{(N_{\textrm{X}}/N_{\textrm{Y}})_\odot}$. The
solar abundance set is that of Ref.~\cite{2009ARAA..47..481A}, with a
solar metallicity Z$_\odot$ = 0.0134.

\subsection{Star-forming environment of CEMP stars}

A great deal of effort has been directed in the literature towards
understanding theoretically, the origin of the most metal-poor stars,
and in particular, the large fraction that is C-rich. These efforts
have been further perturbed by the fact that CEMP stars do not form a
homogenous group, but can rather be further subdivided into two main
populations~\cite{2005ARAA..43..531B}: carbon-rich stars that show
an excess of heavy neutron-capture elements (CEMP-s, CEMP-r, and
CEMP-r/s), and carbon-rich stars with a normal pattern of the heavy
elements (CEMP-no). In the following sections, we focus on stars with
[Fe/H] $\leq$ --3.0, which have been shown to fall almost exclusively
in the CEMP-no subset~\cite{2010IAUS..265..111A}.

A number of theoretical scenarios have been proposed to explain the
observed elemental abundances of these stars, though there is no
universally accepted hypothesis.  The most extensively studied
mechanism to explain the origin of CEMP-no stars is the mixing and
fallback model, where a ``faint" Population III SN explodes, but due
to a relatively low explosion energy, only ejects its outer layers,
rich in lighter elements (up to magnesium); its innermost layers, rich
in iron and heavier elements, fall back onto the remnant and are not
recycled in the ISM~\cite{2003Natur.422..871U,2005ApJ...619..427U}.
This potential link between primeval SNe and CEMP-no stars is
supported by recent studies which demonstrate that the observed ratio
of carbon-enriched to carbon-normal stars with [Fe/H] $<$ --3.0 is
accurately reproduced if SNe were the main source of metal-enrichment
in the early Universe~\cite{2014MNRAS.445.3039D,2014ApJ...791..116C}.
Furthermore, the observed abundance patterns of CEMP-no stars have
been found to be generally well matched by the nucleosynthetic yields
of primordial faint SNe
\cite{2005ApJ...619..427U,2005Sci...309..451I,2009ApJ...693.1780J,2013ApJ...762...27Y,2014Natur.506..463K,2014ApJ...792L..32I,2014ApJ...794..100M,2015MNRAS.454.4250M,2014ApJ...785...98T,2015AA...579A..28B}. These
findings suggest that most of the CEMP-no stars were probably born out
of gas enriched by massive, first-generation stars that ended their
lives as Type II SNe with low levels of mixing and a high degree of
fallback.

Under such circumstances, the gas clouds which collapse and fragment
to form these CEMP-no stars and their protostellar disks may contain
significant amounts of carbon dust grains.  Observationally, dust
formation in SNe ejecta has been inferred from isotopic anomalies in
meteorites where graphite, SiC, and Si$_3$N$_4$ dust grains have been
identified as SNe condensates
\cite{1998M&PS...33..549Z}. Furthermore, in situ dust formation has
been unambiguously detected in the expanding ejecta of SNe such as SN
1987A~\cite{1989LNP...350..164L, 2014ApJ...782L...2I} and SN 1999em
\cite{2003MNRAS.338..939E}. The existence of cold dust has also been
verified in the supernova remnant of Cassiopeia A by SCUBA's recent
submillimeter observations, and a few solar masses worth of dust is
estimated to have condensed in the ejecta~\cite{2003Natur.424..285D}.

Theoretical calculations of dust formation in primordial
core-collapsing SNe have demonstrated the condensation of a variety of
grain species, starting with carbon, in the ejecta, where the mass
fraction tied up in dust grains grows with increasing progenitor mass
\cite{1989ApJ...344..325K,2001MNRAS.325..726T,2003ApJ...598..785N}. Ref.~\cite{2014ApJ...794..100M,2015MNRAS.454.4250M}
consider, in particular, dust formation in weak Population III SNe
ejecta, the type believed to have polluted the birth clouds of CEMP-no
stars.  Tailoring the SN explosion models to reproduce the observed
elemental abundances of CEMP-no stars, they find that: (i) for all the
progenitor models investigated, amorphous carbon (AC) is the only
grain species that forms in significant amounts; this is a consequence
of extensive fallback, which results in a distinct, carbon-dominated
ejecta composition with negligible amounts of other metals, such as
Mg, Si, and Al, that can enable the condensation of alternative grain
types; (ii) the mass of carbon locked into AC grains increases when
the ejecta composition is characterized by an initial mass of C
greater than the O mass; this is particularly true in zero metallicity
supernova progenitors, which undergo less mixing than their solar
metallicity counterparts~\cite{2009ApJ...693.1780J}; in their
stratified ejecta, C-grains are found only to form in layers where C/O
$>$ 1; in layers where C/O $<$ 1, all the carbon is promptly locked in
CO molecules; (iii) depending on the model, the mass fraction of dust
(formed in SNe ejecta) that survives the passage of a SN reverse shock
ranges between 1 to 85\%; this fraction is referred to as the carbon
condensation efficiency; (iv) further grain growth in the collapsing
birth clouds of CEMP-no stars, due to the accretion of carbon atoms in
the gas phase onto the remaining grains, occurs only if C/O $>$ 1 and
is otherwise hindered by the formation of CO molecules.

Besides the accumulation of carbon-rich grains imported from the SNe
ejecta, Fischer-Trope-type reactions (FTTs) may also contribute to
solid carbon enrichment in the protostellar disks of CEMP-no stars by
enabling the conversion of nebular CO and H$_2$ to other forms of
carbon~\cite{2001M&PS...36...75K}. Furthermore,
in carbon-rich gas, the equilibrium condensation sequence changes
signifcantly from the sequence followed in solar composition gas where
metal oxides condense first. In nebular gas with C/O $\gtrsim$ 1,
carbon-rich compounds such as graphite, carbides, nitrides, and
sulfides are the highest temperature condensates ($T \approx$
1200-1600 K)~\cite{1975GeCoA..39..389L}. Thus, if planet formation is
to proceed in this C-rich gas, the protoplanetary disks of these
CEMP-no stars may spawn many carbon planets.

\subsection{Orbital Radii of Potential Carbon Planets }

Given the significant abundance of carbon grains, both imported from
SNe ejecta and produced by equilibrium and non-equilibrium mechanisms
operating in the C-rich protoplanetary disks, the emerging question
is: would these dust grains have enough time to potentially coagulate
and form planets around their host CEMP-no stars?

In the core accretion model, terrestrial planet formation is a
multi-step process, starting with the aggregation and settling of dust
grains in the protoplanetary disk
\cite{1993ARAA..31..129L,2000prpl.conf..533B,2006RPPh...69..119P,2007prpl.conf..639N,Armitagebook,2011ApJ...736...89J}. In
this early stage, high densities in the disk allow particles to grow
from submicron-size to meter-size through a variety of collisional
processes including Brownian motion, settling, turbulence, and radial
migration. The continual growth of such aggregates by coagulation and
sticking eventually leads to the formation of kilometer-sized
planetesimals, which then begin to interact gravitationally and grow
by pairwise collisions, and later by runaway growth
\cite{1993ARAA..31..129L}.  In order
for terrestrial planets to ultimately form, these processes must all
occur within the lifetime of the disk itself, a limit which is set by
the relevant timescale of the physical phenomena that drive disk
dissipation.

A recent study by Ref.~\cite{2009ApJ...705...54Y} of clusters in the
Extreme Outer Galaxy (EOG) provides observational evidence that
low-metallicity disks have shorter lifetimes ($<$ 1 Myr) compared to
solar metallicity disks ($\sim$ 5-6 Myr). This finding is consistent
with models in which photoevaporation by energetic (ultraviolet or
X-ray) radiation of the central star is the dominant disk dispersal
mechanism. While the opacity source for EUV (extreme-ultraviolet)
photons is predominantly hydrogen and is thus metallicity-independent,
X-ray photons are primarily absorbed by heavier elements, mainly
carbon and oxygen, in the inner gas and dust shells. Therefore, in low
metallicity environments where these heavy elements are not abundant
and the opacity is reduced, high density gas at larger columns can be
ionized and will experience a photoevaporative flow if heated to high
enough temperatures~\cite{2009ApJ...690.1539G,2010MNRAS.402.2735E}.

Assuming that photoevaporation is the dominant mechanism through which
circumstellar disks lose mass and eventually dissipate, we adopt the
metallicity-dependent disk lifetime, derived in
Ref.~\cite{2010MNRAS.402.2735E} using X-ray+EUV models
\cite{2009ApJ...699.1639E},
\begin{equation}
t_{disk}\propto Z^{0.77(4-2p)/(5-2p)}
\end{equation}
where $Z$ is the total metallicity of the disk and $p$ is the
power-law index of the disk surface density profile ($\Sigma \propto
r^{-p}$). A mean power-law exponent of $p \sim$ 0.9 is derived by
modeling the spatially resolved emission morphology of young stars at
(sub)millimeter wavelengths
\cite{2009ApJ...700.1502A,2010ApJ...723.1241A} and the timescale is
normalized such that the mean lifetime for disks of solar metallicity
is 2 Myr~\cite{2010MNRAS.402.2735E}.  thus the disk lifetime, is
dominated by carbon dust grains in the CEMP-no stars considered here,
we adopt the carbon abundance relative to solar [C/H] as a proxy for
the overall metallicity $Z$ We adopt the carbon abundance relative to
solar [C/H] as a proxy for the overall metallicity $Z$ since the
opacity, which largely determines the photoevaporation rate, and thus
the disk lifetime, is dominated by carbon dust grains in the CEMP-no
stars we consider in this work.

The timescale for planet formation is believed to be effectively set
by the time it takes dust grains to settle into the disk midplane. The
subsequent process of runaway planetesimal formation, possibly
occurring via a series of pairwise collisions, must be quick, since
otherwise, the majority of the solid disk material would radially
drift towards the host star and evaporate in the hot inner regions of
the circumstellar disk~\cite{Armitagebook}.  We adopt the
one-particle model of Ref.~\cite{2005AA...434..971D} to follow the mass
growth of dust grains via collisions as they fall through and sweep up
the small grains suspended in the disk.  Balancing the gravitational
force felt by a small dust particle at height $z$ above the mid-plane
of a disk with the aerodynamic drag (in the Epstein regime) gives a
dust settling velocity of
\begin{equation}
v_{sett}=\frac{dz}{dt}=\frac{3\Omega_K^2zm}{4\rho c_s\sigma_d}
\end{equation}
where $\sigma_d=\pi a^2$ is the cross-section of the dust grain with
radius $a$ and $c_s=\sqrt{k_BT(r)/\mu m_H}$ is the isothermal sound
speed with $m_H$ being the mass of a hydrogen atom and $\mu$=1.36
being the mean molecular weight of the gas (including the contribution
of helium). $\Omega_K=\sqrt{GM_*/r^3}$ is the Keplerian velocity of
the disk at a distance $r$ from the central star of mass $M_*$, which
we take to be $M_*$ = 0.8 M$_\odot$ as representative of the
low-masses associated with CEMP-no stars
\cite{2002Natur.419..904C,2015ARAA..53..631F}.  The disk is assumed
to be in hydrostatic equilibrium with a density given by
\begin{equation}
\rho(z,r)=\frac{\Sigma(r)}{h\sqrt{2\pi}}\exp{\left(-\frac{z^2}{2h^2}\right)}
\end{equation}
where the disk scale height is $h=c_s/\Omega_k$. For the disk surface
density $\Sigma(r)$ and temperature $T(r)$ profiles, we adopt the
radial power-law distributions fitted to (sub-)millimeter observations
of circumstellar disks around young stellar objects
\cite{2005ApJ...631.1134A,2009ApJ...700.1502A,2010ApJ...723.1241A},
\begin{equation}
T(r)=200 \textrm{ K} \left(\frac{r}{1 \textrm{ AU}}\right)^{-0.6} 
\end{equation}
\begin{equation}
\Sigma(r)=10^3 \textrm{ g/cm$^2$}\left(\frac{r}{1 \textrm{ AU}}\right)^{-0.9} \,\,.
\end{equation}
Although these relations were observationally inferred from disks with
solar-like abundances, we choose to rely on them for our purposes
given the lack of corresponding measurements for disks around stars
with different abundance patterns.

The rate of grain growth, $dm/dt$, is determined by the rate at which
grains, subject to small-scale Brownian motion, collide and stick
together as they drift towards the disk mid-plane through a sea of
smaller solid particles. If coagulation results from every collision,
then the mass growth rate of a particle is effectively the amount of
solid material in the volume swept out the particle's geometric
cross-section,
\begin{equation}
\frac{dm}{dt}=f_{dg}\rho\sigma_d\left(v_{rel}+\frac{dz}{dt}\right)
\end{equation}
where $dz/dt$ is the dust settling velocity given by equation (2) and
\begin{equation}
v_{rel}=\sqrt{\frac{8k_BT(m_1+m_2)}{\pi m_1m_2}}\approx\sqrt{\frac{8k_BT}{\pi m}}
\end{equation}
is the relative velocity in the Brownian motion regime between grains
with masses $m_1=m_2=m$. To calculate the dust-to-gas mass ratio in
the disk $f_{dg}$, we follow the approach in
Ref.~\cite{2014ApJ...782...95J} and relate two expressions for the mass
fraction of C: (i) the fraction of carbon in the dust,
$f_{dg}M_{C,dust}/M_{dust}$, where $M_{dust}$ is the total dust mass
and $M_{C,dust}$ is the carbon dust mass ; and (ii) the fraction of
carbon in the gas, $\mu_C n_C/\mu n_H$, where $\mu_C$ is the molecular
weight of carbon ($\sim12m_p$) and $n_C$ and $n_H$ are the carbon and
hydrogen number densities, respectively.

\begin{table}
\centering\begin{minipage}{90mm}
  \caption{Basic data$^a$ for CEMP stars considered in this section}
  \begin{tabular}{@{}lccccccccc@{}}
  \hline
  \hline
 Star & $\log{g}^b$ & [Fe/H] & [C/Fe] & C/O$^c$ & Source$^d$\\
 \hline
  HE\,0107-5240 & 2.2 & -5.44 & 3.82  & 14.1 & 1,2\\
 SDSS J0212+0137 & 4.0 & -3.57 & 2.26 & 2.6 & 3\\
 SDSS J1742+2531 & 4.0 & -4.77 & 3.60 & 2.2 & 3\\
 G\,77-61 & 5.1 & -4.03 & 3.35 & 12.0 & 4, 5\\
 HE\,2356-0410$^e$ & 2.65 & -3.19 & 2.61 & $>$14.1 & 6\\
\hline
\hline
\end{tabular}
\\ $^a$ Abundances based on one-dimensional LTE model-atmosphere
analyses\\ $^b$ Logarithm of the gravitational acceleration at the
surface of stars expressed in cm\,s$^{-2}$\\ $^c$ C/O =
$N_{\textrm{C}}/N_{\textrm{O}} =
10^{\textrm{[C/O]}+\log{\epsilon(\textrm{C})_\odot}-\log{\epsilon(\textrm{O})_\odot}}$\\ $^d$
\textbf{References:} (1) ~\cite{2004ApJ...603..708C}; (2)
~\cite{2006ApJ...644L.121C}; (3) ~\cite{2015AA...579A..28B}; (4)
~\cite{2005AA...434.1117P}; (5) ~\cite{2007AJ....133.1193B}; (6)
~\cite{2014AJ....147..136R}.\\ $^e$ CS\,22957-027
\end{minipage}
\end{table}

\noindent We then assume that a fraction $f_{cond}$ (referred to from
now on as the carbon condensation efficiency) of all the carbon
present in the gas cloud is locked up in dust, such that
\begin{equation}
f_{cond}\frac{\mu_C n_C}{\mu n_H}=f_{dg}\frac{M_{C,dust}}{M_{dust}} \,\,.
\end{equation}
Since faint Population III SNe are believed to have polluted the birth
clouds of CEMP-no stars, and the only grain species that forms in
non-negligible amounts in these ejecta is amorphous carbon
\cite{2014ApJ...794..100M,2015MNRAS.454.4250M}, we set
$M_{dust}=M_{C,dust}$.  Rewriting equation (8) in terms of abundances
relative to the Sun, we obtain
\begin{equation}
f_{dg}=f_{cond}\frac{\mu_C}{\mu}10^{\textrm{[C/H]}+\log{\epsilon(C)_\odot}-12}
\end{equation}
where $\log{\epsilon(C)_\odot}$=8.43$\pm$0.05
\cite{2009ARAA..47..481A} is the solar carbon abundance.

For a specified metallicity [C/H] and radial distance $r$ from the
central star, we can then estimate the time it takes for dust grains
to settle in the disk by integrating equations (2) and (6) from an
initial height of $z(t=0)=4h$ with an initial dust grain mass of
$m(t=0)=4\pi a_{init}^3\rho_d/3$. The specific weight of dust is set
to $\rho_d$=2.28 g cm$^{-3}$, reflecting the material density of
carbon grains expected to dominate the circumstellar disks of CEMP-no
stars. The initial grain size $a_{init}$ is varied between 0.01 and 1
$\mu$m to reflect the range of characteristic radii of carbon grains
found when modeling CEMP-no star abundance patterns
\cite{2014ApJ...794..100M}. Comparing the resulting dust-settling
timescale to the disk lifetime given by equation (1) for the specified
metallicity, we can then determine whether there is enough time for
carbon dust grains to settle in the mid-plane of the disk and there
undergo runaway planetesimal formation before the disk is dissipated
by photoevaporation. For the purposes of this simple model, we
neglected possible turbulence in the disk which may counteract the
effects of vertical settling, propelling particles to higher altitudes
and thus preventing them from fully settling into the disk mid-plane
\cite{Armitagebook}. We have also not accounted for the effects of
radial drift, which may result in the evaporation of solid material in
the hot inner regions of the circumstellar disk.

\begin{figure}
\label{fig1}
\hspace{-1.3cm}\includegraphics[width=105mm,height=80mm]{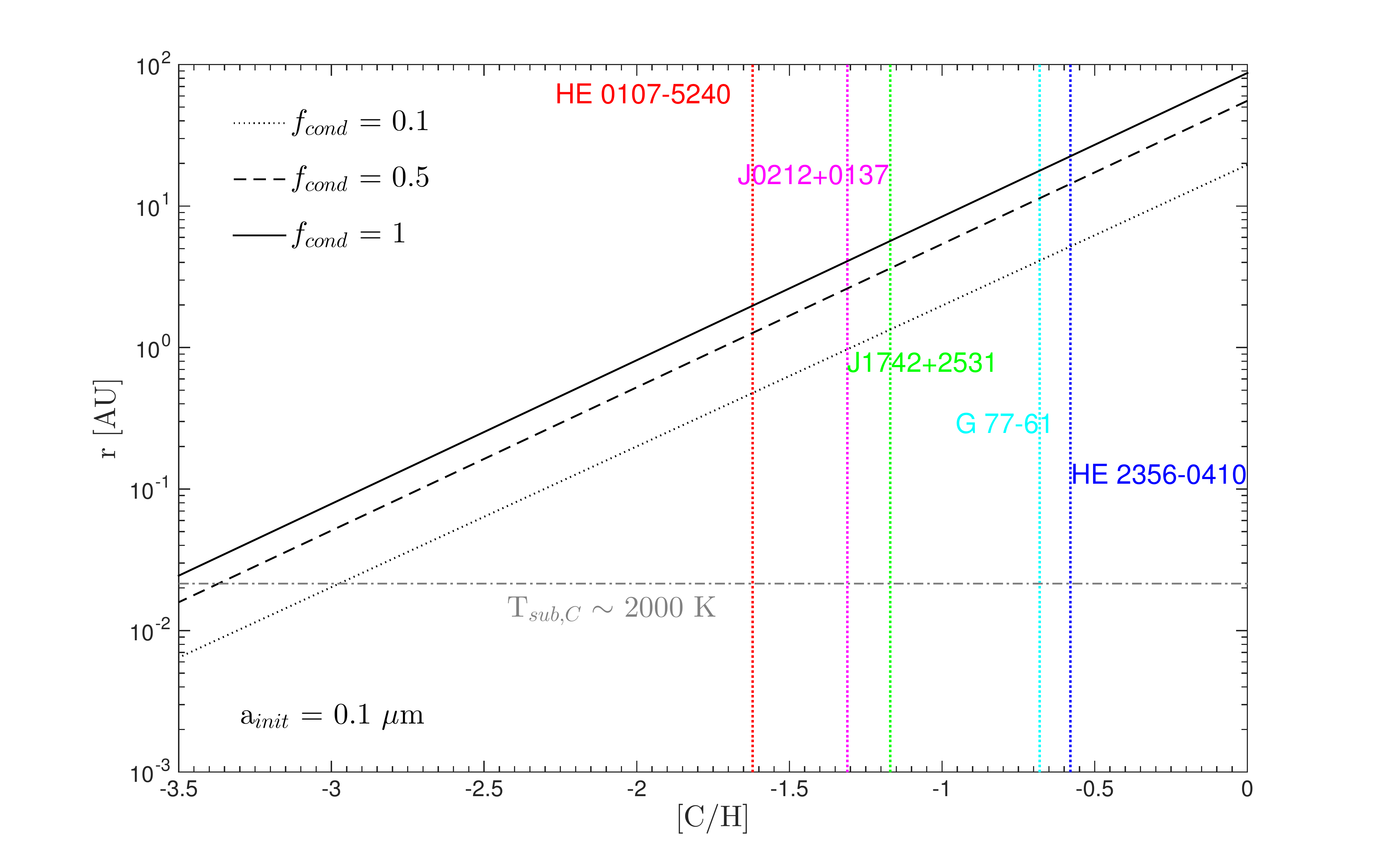}
\caption{The maximum distance $r_{max}$ from the host star out to
  which planetesimal formation is possible as a function of the star's
  metallicity, expressed as the carbon abundance relative to that of
  the Sun, [C/H]. The dotted, dashed, and solid black curves
  correspond to the results obtained assuming carbon condensation
  efficiencies of 10\%, 50\%, and 100\%, respectively, and an initial
  grain size of $a_{init}$ = 0.1 $\mu$m. The gray dash-dotted curve
  corresponds to the distance at which the disk temperature approaches
  the sublimation temperature of carbon dust grains,
  $T_{sub,\textrm{C}} \sim$ 2000 K; the formation of carbon
  planetesimals will therefore be suppressed at distances that fall
  below this line, $r \lesssim$ 0.02 AU. The colored vertical lines
  represent various observed CEMP stars with measured carbon
  abundances, [C/H].}
\end{figure}

As the dust settling timescale is dependent on the disk surface
density $\Sigma(r)$ and temperature $T(r)$, we find that for a given
metallicity, [C/H], there is a maximum distance $r_{max}$ from the
central star out to which planetesimal formation is possible. At
larger distances from the host star, the dust settling timescale
exceeds the disk lifetime and so carbon planets with semi-major axes
$r > r_{max}$ are not expected to form.  A plot of the maximum
semi-major axis expected for planet formation around a CEMP-no star as
a function of the carbon abundance relative to the Sun [C/H] is shown
in Figure \ref{fig1} for carbon condensation efficiencies ranging
between $f_{cond}$ = 0.1 and 1. As discovered in
Ref.~\cite{Johnson} where the critical iron abundance for
terrestrial planet formation is considered as a function of the
distance from the host star, we find a linear relation between [C/H]
and $r_{max}$,
\begin{equation}
\textrm{[C/H]}=\log{\left(\frac{r_{max}}{1 \textrm{ AU}}\right)}-\alpha
\end{equation}
where $\alpha$ = 1.3, 1.7, and 1.9 for $f_{cond}$ = 0.1, 0.5, and 1,
respectively, assuming an initial grain size of $a_{init}$ = 0.1
$\mu$m. These values for $\alpha$ change by less than 1\% for smaller
initial grain sizes, $a_{init}$ = 0.01 $\mu$m, and by no more than 5\%
for larger initial grain sizes $a_{init}$ = 1 $\mu$m; given this weak
dependence on $a_{init}$, we only show our results for a single
initial grain size of $a_{init}$ = 0.1 $\mu$m. The distance from the
host star at which the temperature of the disk approaches the
sublimation temperature of carbon dust, $T_{sub,C} \sim$ 2000 K
\cite{2011EP&S...63.1067K}, is depicted as well (dash-dotted gray
curve). At distances closer to the central star than $r \simeq$ 0.02
AU, temperatures well exceed the sublimation temperature of carbon
grains; grain growth and subsequent carbon planetesimal formation are
therefore quenched in this inner region.

Figure \ref{fig1} shows lines representing various observed CEMP stars with
measured carbon abundances, mainly, HE\,0107-5240
\cite{2002Natur.419..904C,2004ApJ...603..708C}, SDSS J0212+0137
\cite{2015AA...579A..28B}, SDSS J1742+2531
\cite{2015AA...579A..28B}, G\,77-61
\cite{1977ApJ...216..757D,2005AA...434.1117P,2007AJ....133.1193B},
and HE\,2356-0410
\cite{1997ApJ...489L.169N,2014AJ....147..136R}. These stars all have
iron abundances (relative to solar) [Fe/H] $<$ -3.0, carbon abundances
(relative to solar) [C/Fe] $>$ 2.0, and carbon-to-oxygen ratios C/O
$>$ 1. This latter criteria maximizes the abundance of solid carbon
available for planet formation in the circumstellar disks by
optimizing carbon grain growth both in stratified SNe ejecta and
later, in the collapsing molecular birth clouds of these stars. It
also advances the possibility of carbon planet formation by ensuring
that planet formation proceeds by a carbon-rich condensation sequence
in the protoplanetary disk.  SDSS J0212+0137 and HE\,2356-0410 have
both been classified as CEMP-no stars, with measured barium abundances
[Ba/Fe] $<$ 0 (as defined in ~\cite{2005ARAA..43..531B}); the other
three stars are Ba-indeterminate, with only high upper limits on
[Ba/Fe], but are believed to belong to the CEMP-no subclass given
their light-element abundance patterns.  The carbon abundance, [C/H],
dominates the total metal content of the stellar atmosphere in these
five CEMP objects, contributing more than 60\% of the total
metallicity in these stars.  A summary of the relevant properties of
the CEMP stars considered in this analysis can be found in Table 1.
We find that carbon planets may be orbiting iron-deficient stars with
carbon abundances [C/H] $\sim$ -0.6, such as HE\,2356-0410, as far out
as $\sim$ 20 AU from their host star in the case where $f_{cond}$ =
1. Planets forming around stars with less carbon enhancement,
i.e. HE\,0107-5240 with [C/H] $\sim$ -1.6, are expected to have more
compact orbits, with semi-major axes $r <$ 2 AU. If the carbon
condensation efficiency is only 10\%, the expected orbits grow even
more compact, with maximum semi-major axes of $\sim$ 5 and 0.5 AU,
respectively.

\subsection{Mass-Radius Relationship for Carbon Planets}

Next we present the relationship between the mass and radius of carbon
planets that we have shown may theoretically form around CEMP-no
stars. These mass-radius relations have already been derived in the
literature for a wide range of rocky and icy exoplanet compositions
\cite{1969ApJ...158..809Z,2004Icar..169..499L,2006Icar..181..545V,2007ApJ...659.1661F,2007ApJ...669.1279S}. Here,
we follow the approach of Ref.~\cite{1969ApJ...158..809Z} and solve the
three canonical equations of internal structure for solid planets,
\begin{enumerate}
\item mass conservation
\begin{equation}
\frac{dm(r)}{dr}=4\pi r^2 \rho(r) \,,
\end{equation}
\item hydrostatic equilbrium
\begin{equation}
\frac{dP(r)}{dr}=-\frac{Gm(r)\rho(r)}{r^2} \,,\,\,\textrm{and}
\end{equation}
\item the equation of state (EOS)
\begin{equation}
P(r) = f\left(\rho(r),T(r)\right) \,,
\end{equation}
\end{enumerate}
where $m(r)$ is the mass contained within radius $r$, $P(r)$ is the
pressure, $\rho(r)$ is the density of the spherical planet, and $f$ is
the unique equation of state (EOS) of the material of interest, in
this case, carbon.

Carbon grains in circumstellar disks most likely experience many shock
events during planetesimal formation which may result in the
modification of their structure. The coagulation of dust into clumps,
the fragmentation of the disk into clusters of dust clumps, the
merging of these clusters into $\sim$ 1 km planetesimals, the
collision of planetesimals during the accretion of meteorite parent
bodies, and the subsequent collision of the parent bodies after their
formation all induce strong shock waves that are expected to
chemically and physically alter the materials.
Subject to these high temperatures and pressures, the amorphous carbon
grains polluting the protoplanetary disks around CEMP stars are
expected to undergo graphitization and may even crystallize into
diamond~\cite{1987ApJ...319L.109T,1996AA...315..222P,Takai2003}.
In our calculations, the equation of state at low pressures, $P \leq$
14 GPa, is set to the third-order finite strain Birch-Murnagham EOS
(BME; ~\cite{1947PhRv...71..809B}) for graphite,
\begin{equation}
P=\frac{3}{2}K_0\left(\eta^{7/3}-\eta^{5/3}\right)\left[1+\frac{3}{4}
\left(K_{0}^{'} - 4\right)\left(\eta^{2/3}-1\right)\right]
\end{equation}
where $\eta=\rho/\rho_0$ is the compression ratio with respect to the
ambient density, $\rho_0$, $K_0$ is the bulk modulus of the material,
and $K_{0}^{'}$ is the pressure derivative. Empirical fits to experimental
data yields a BME EOS of graphite ($\rho_0$ = 2.25 g cm$^{-3}$) with
parameters $K_0$ = 33.8 GPa and $K_{0}^{'}$ = 8.9
\cite{1989PhRvB..3912598H}. At 14 GPa, we incorporate the phase
transition from graphite to diamond
\cite{1976Natur.259...38N,1989PhRvB..3912598H} and adopt the Vinet
EOS~\cite{1987JGR....92.9319V,1989JPCM....1.1941V},
\begin{equation}
P=3K_0\eta^{2/3}\left(1-\eta^{-1/3}\right)\exp{\left[\frac{3}{2}\left(K_{0}^{'}-1\right)\left(1-\eta^{-1/3}\right)\right]}
\end{equation}
with $K_0$ = 444.5 GPa and $K_{0}^{'}$ = 4.18 empirically fit for diamond,
$\rho_0$ = 3.51 g cm$^{-3}$~\cite{2008PhRvB..77i4106D}. (As pointed
out in Ref.~\cite{2007ApJ...669.1279S}, the BME EOS is not fit to be
extrapolated to high pressures since it is derived by expanding the
elastic potential energy as a function of pressure keeping only the
lowest order terms.) Finally, at pressures $P \gtrsim$ 1300 GPa where
electron degeneracy becomes increasingly important, we use the
Thomas-Fermi-Dirac (TFD) theoretical EOS
(~\cite{1967PhRv..158..876S}; equations (40)-(49)), which intersects
the diamond EOS at $P \sim$ 1300 GPa.  Given that the full
temperature-dependent carbon EOSs are either undetermined or dubious
at best, all three EOSs adopted in this section are room-temperature EOSs
for the sake of practical simplification.

\begin{figure}
\label{fig2}
\hspace{-.7cm}\includegraphics[width=105mm,height=80mm]{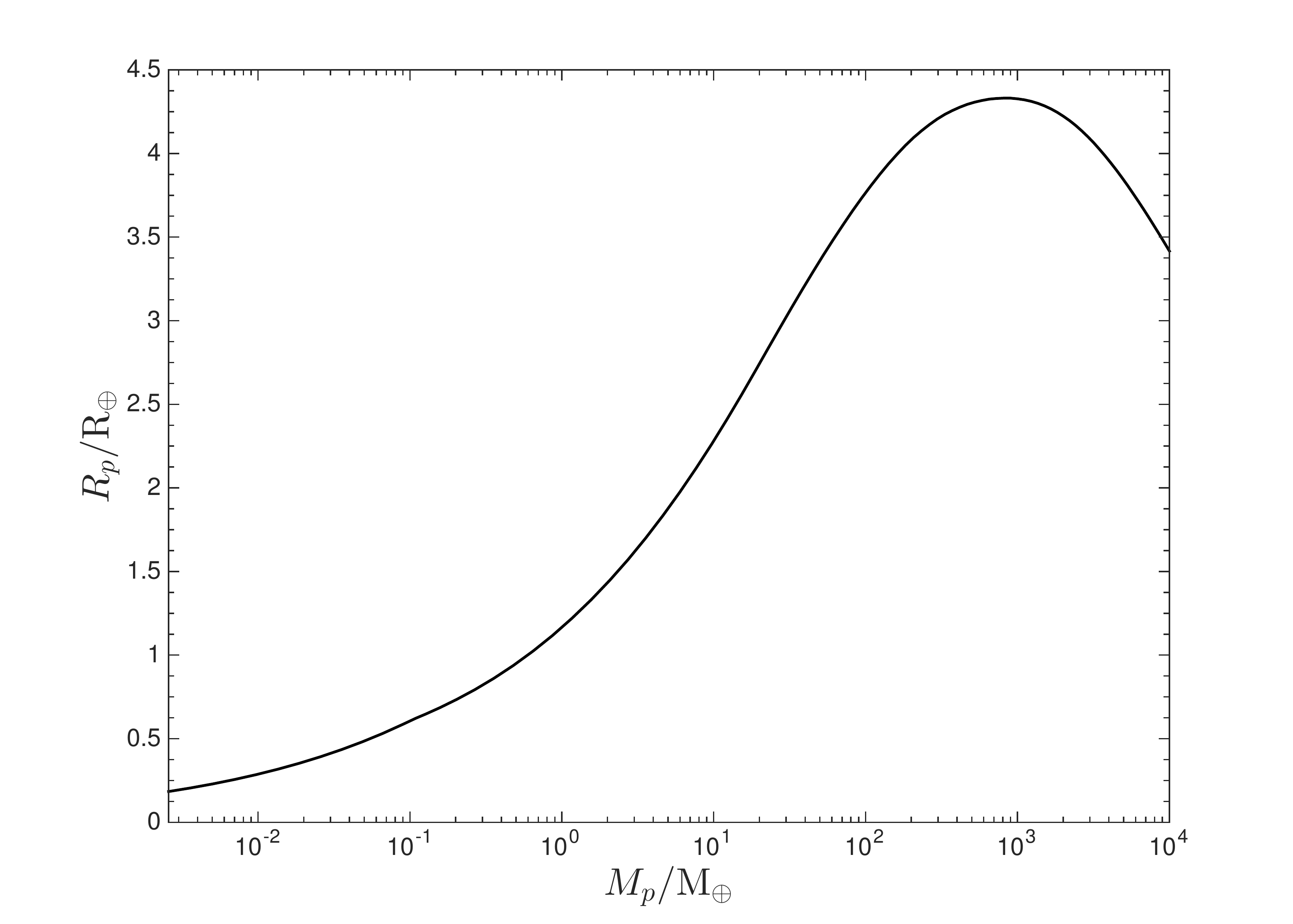}
\caption{Mass-radius relation for solid homogenous, pure carbon planet}
\end{figure}

Using a fourth-order Runge-Kutta scheme, we solve the system of
equations simultaneously, numerically integrating equations (11) and
(12) begining at the planet's center with the inner boundary
conditions $M(r=0)$ = 0 and $P(r=0)$ = $P_{\textrm{central}}$, where
$P_{\textrm{central}}$ is the central pressure. The outer boundary
condition $P(r=R_p)$ = 0 then defines the planetary radius $R_p$ and
total planetary mass $M_p = m(r=R_p)$. Integrating these equations for
a range of $P_{\textrm{central}}$, with the appropriate EOS,
$P=P(\rho)$, to close the system of equations, yields the mass-radius
relationship for a given composition. We show this mass-radius
relation for a purely solid carbon planet in Figure \ref{fig2}. We
find that for masses $M_p \lesssim$ 800 M$_\oplus$, gravitational
forces are small compared with electrostatic Coulomb forces in
hydrostatic equilibrium and so the planet's radius increases with
increasing mass, $R_p \propto M_p^{1/3}$. However, at larger masses,
the electrons are pressure-ionized and the resulting degeneracy
pressure becomes significant, causing the planet radius to become
constant and even decrease for increasing mass, $R_p \propto
M_p^{-1/3}$~\cite{Hubbardbook}.  Planets which fall within the mass
range 500 $\lesssim M_p \lesssim$ 1300 M$_\oplus$, where the competing
effects of Coulomb forces and electron degeneracy pressure cancel each
other out, are expected to be approximately the same size, with $R_p
\simeq$ 4.3 R$_\oplus$, the maximum radius of a solid carbon
planet. (In the case of gas giants, the planet radius can increase due
to accretion of hydrogen and helium.)

Although the mass-radius relation illustrated in Figure \ref{fig2} may
alone not be enough to confidently distinguish a carbon planet from a
water or silicate planet, the unique spectral features in the
atmospheres of these carbon planets may provide the needed
fingerprints. At high temperatures ($T \gtrsim$ 1000 K), the
absorption spectra of massive ($M \sim$ 10 - 60 M$_\oplus$) carbon
planets are expected to be dominated by CO, in contrast with the
H$_2$O-dominated spectra of hot massive planets with solar-composition
atmospheres~\cite{2005astro.ph..4214K}.  The atmospheres of low-mass
($M \lesssim$ 10 M$_\oplus$) carbon planets are also expected to be
differentiable from their solar-composition counterparts due to their
abundance of CO and CH$_4$, and lack of oxygen-rich gases like CO$_2$,
O$_2$, and O$_3$~\cite{2005astro.ph..4214K}.  Furthermore, carbon
planets of all masses at low temperatures are expected to accommodate
hydrocarbon synthesis in their atmospheres; stable long-chain
hydrocarbons are therefore another signature feature that can help
observers distinguish the atmospheres of cold carbon planets and more
confidently determine the bulk composition of a detected planet
\cite{2005astro.ph..4214K}.
 
The detection of theoretically proposed carbon planets around CEMP
stars will provide us with significant clues regarding how early
planet formation may have started in the Universe.  While direct
detection of these extrasolar planets remains difficult given the low
luminosity of most planets, techniques such as the transit method are
often employed to indirectly spot exoplanets and determine physical
parameters of the planetary system.  When a planet ``transits" in
front of its host star, it partially occludes the star and causes its'
observed brightness to drop by a minute amount. If the host star is
observed during one of these transits, the resulting dip in its
measured light curve can yield information regarding the relevant
sizes of the star and the planet, the orbital semi-major axis, and the
orbital inclination, among other characterizing properties.

\subsection{Section Summary and Implications}
We explored the possibility of carbon planet formation around the
iron-deficient, carbon-rich subset of low-mass stars known as CEMP
stars. The observed abundance patterns of CEMP-no stars suggest that
these stellar objects were probably born out of gas enriched by
massive first-generation stars that ended their lives as Type II SNe
with low levels of mixing and a high degree of fallback.  The
formation of dust grains in the ejecta of these primordial
core-collapsing SNe progenitors has been observationally confirmed and
theoretically studied. In particular, amorphous carbon is the only
grain species found to condense and form in non-negligible amounts in
SN explosion models that are tailored to reproduce the abundance
patterns measured in CEMP-no stars.  Under such circumstances, the gas
clouds which collapse and fragment to form CEMP-no stars and their
protoplanetary disks may contain significant amounts of carbon dust
grains imported from SNe ejecta.  The enrichment of solid carbon in
the protoplanetary disks of CEMP stars may then be further enhanced by
Fischer-Trope-type reactions and carbon-rich condensation sequences,
where the latter occurs specifically in nebular gas with C/O $\gtrsim$
1.

For a given metallicity [C/H] of the host CEMP star, the maximum
distance out to which planetesimal formation is possible can then be
determined by comparing the dust-settling timescale in the
protostellar disk to the expected disk lifetime. Assuming that disk
dissipation is driven by a metallicity-dependent photoevaporation
rate, we find a linear relation between [C/H] and the maximum
semi-major axis of a carbon planet orbiting its host CEMP star.  Very
carbon-rich CEMP stars, such as G\,77-61 and HE\,2356-0410 with [C/H]
$\simeq$ -0.7 -- -0.6, can host carbon planets with semi-major axes as
large $\sim$ 20 AU for 100\% carbon condensation efficiencies; this
maximum orbital distance reduces to $\sim$ 5 AU when the condensation
efficiency drops by an order of magnitude.  In the case of the
observed CEMP-no stars HE\,0107-5240, SDSS\,J0212+0137, and
SDSS\,J1742+2531, where the carbon abundances are in the range [C/H]
$\simeq$ -1.6 -- -1.2, we expect more compact orbits, with maximum
orbital distances $r_{max} \simeq$ 2, 4, and 6 AU, respectively, for
$f_{cond}$ = 1 and $r_{max} \simeq$ 0.5 - 1 AU for $f_{cond}$ = 0.1.

%We then use the linear relation found between [C/H] and $r_{max}$,
%along with the theoretical mass-radius relation derived for a solid,
%pure carbon planet (\S4), to compute the three observable
%characteristics of planetary transits: the orbital period, the transit
%depth, and the transit duration. We find that the relative change in
%flux, $\Delta F$, caused by an Earth-mass carbon planet transiting
%across its host CEMP star ranges from $\sim$ 0.0001\% for a stellar
%radius of $R_* \sim$ 10 R$_\odot$ to $\sim$ 0.01\% for a solar-sized
%stellar host.  
While the shallow transit depths of Earth-mass carbon
planets around HE\,0107-5240 and HE\,2356-0410 may evade detection,
current and future space-based transit surveys promise to achieve the
precision levels ($\sim$ 0.001\%) necessary to detect
planetary systems around CEMP stars such as SDSS\,J0212+0137,
SDSS\,J1742+2531, and G\,77-61. If gas giant (Jupiter scale) planets
form around CEMP stars, their transits would be much easier to detect
than rocky planets. However, they are not likely to host life as we
know it. There are a number of ongoing, planned, and proposed space
missions committed to this cause, including CoRot (COnvection ROtation
and planetary Transits), Kepler, PLATO (PLAnetary Transits and
Oscillations of stars), TESS (Transiting Exoplanet Survey Satellite),
and ASTrO (All Sky Transit Observer), which are expected to achieve
precisions as low as 20-30 ppm (parts per million)
\cite{2009IAUS..253..319B,2015ARAA..53..409W}. 

Short orbital periods and long transit durations are also key
ingredients in boosting the probability of transit detection by
observers. G\,77-61 is not an optimal candidate in these respects
since given its large carbon abundance ([C/Fe] $\sim$ 3.4), carbon
planets may form out to very large distances and take up to a century
to complete an orbit around the star for $f_{cond}$ = 1 ($P_{max}
\sim$ 10 years for 10\% carbon condensation efficiency). The small
stellar radius, $R_* \sim$ 0.5 R$_\odot$, also reduces chances of
spotting the transit since the resulting transit duration is only
$\sim$ 30 hours at most. Carbon planets around larger CEMP stars with
an equally carbon-rich protoplanetary disk, such as HE\,2356-0410
($R_* \sim$ 7 R$_\odot$), have a better chance of being spotted, with
transit durations lasting up to $\sim$ 3 weeks. The CEMP-stars
SDSS\,J0212+0137, and SDSS\,J1742+2531 are expected to host carbon
planets with much shorter orbits, $P_{max} \sim$ 16 years for 100\%
condensation efficiency ($P_{max} \sim$ 1 year for $f_{cond}$ = 0.1),
and transit durations that last as long as $\sim$ 60 hours. If the
ability to measure transit depths improves to a precision of 1 ppm,
then potential carbon planets around HE\,0107-5240 are the most likely
to be spotted (among the group of CEMP-no stars considered here),
transiting across the host star at least once every $\sim$ 5 months
(10\% condensation efficiency) with a transit duration of 6 days.

While our calculations place upper bounds on the distance from the
host star out to which carbon planets can form, we note that orbital
migration may alter a planet's location in the circumstellar disk. As
implied by the existence of `hot Jupiters', it is possible for a
protoplanet that forms at radius $r$ to migrate inward either through
gravitational interactions with other protoplanets, resonant
interactions with planetesimals with more compact orbits, or tidal
interactions with gas in the surrounding disk
\cite{2006RPPh...69..119P}.  Since Figure \ref{fig1} only plots
$r_{max}$, the $maximum$ distance out to which a carbon planet with
     [C/H] can form, our results remain consistent in the case of an
     inward migration. However, unless planets migrate inward from
     their place of birth in the disk, we do not expect to find carbon
     exoplanets orbiting closer than $r \simeq$ 0.02 AU from the host
     stars since at such close proximities, temperatures are high
     enough to sublimate carbon dust grains.

Protoplanets can also be gravitationally scattered into wider orbits
through interactions with planetesimals in the
disk\cite{1999AJ....117.3041H,2009ApJ...696.1600V}. Such an outward
migration of carbon planets may result in observations that are
inconsistent with the curves in Figure \ref{fig1}. A planet that
formed at radius $r \ll r_{max}$ still has room to migrate outwards
without violating the `maximum distance' depicted in Figure
\ref{fig1}.
%; however, the outward migration of a carbon planet that
%originally formed at, or near, $r_{max}$ would result in a breach of
%the upper bounds placed on the transiting properties of carbon planets
%(Section 5). In particular, a carbon planet that migrates to a
%semi-major axis $r > r_{max}$ will have an orbital period and a
%transit duration time that exceeds the limits prescribed in
%our equations.

Detection of the carbon planets that we suggest may have formed around
CEMP stars will provide us with significant clues regarding how planet
formation may have started in the early Universe. The formation of
planetary systems not only signifies an increasing degree of
complexity in the young Universe, but it also carries implications for
the development of life at this early junction \cite{Loeb14}. The
lowest metallicity multi-companion system detected to date is around
BD+20 24 57, a K2-giant with [Fe/H] = -1.0 \cite{2009ApJ...707..768N},
a metallicity value once believed to yield low efficiency for planet
formation \cite{2001Icar..152..185G,2005MNRAS.364...29P}. More recent
formulations of the metallicity required for planet formation are
consistent with this observation, estimating that the first Earth-like
planets likely formed around stars with metallicities [Fe/H]
$\lesssim$ -1.0~\cite{Johnson}. The CEMP stars considered in this
section are extremely iron-deficient, with [Fe/H] $\lesssim$ -3.2, and
yet, given the enhanced carbon abundances which dominate the total
metal content in these stars ( [C/H] $\gtrsim$ -1.6), the formation of
solid carbon exoplanets in the protoplanetary disks of CEMP stars
remains a real possibility.

An observational program aimed at searching for carbon planets around
these low-mass Population II stars could therefore potentially shed
light on the question of how early planets, and subsequently, life
could have formed after the Big Bang.

\section{Water Formation During the Epoch of First Metal Enrichment}
\label{waterfirst}

\subsection{Section Background}

Water is an essential ingredient for life as we know it
\cite{Kasting}.  In the interstellar medium (ISM) of the Milky-Way
and also in external galaxies, water has been observed in the gas
phase and as grain surface ice in a wide variety of environments.
These environments include diffuse and dense molecular clouds,
photon-dominated regions (PDRs), shocked gas, protostellar envelopes,
and disks (see review by Ref. \cite{vanDishoeck2014}).

In diffuse and translucent clouds, H$_2$O is formed mainly in gas
phase reactions via ion-molecule sequences~\cite{Herbst1973}.  The
ion-molecule reaction network is driven by cosmic-ray or X-ray
ionization of H and H$_2$, that leads to the formation of H$^+$ and
H$_3^+$ ions. These interact with atomic oxygen and form OH$^+$. A
series of abstractions then lead to the formation of H$_3$O$^+$, which
makes OH and H$_2$O through dissociative recombination.  This
formation mechanism is generally not very efficient, and only a small
fraction of the oxygen is converted into water, the rest remains in
atomic form, or freezes out as water ice~\cite{Hollenbach2009}.

Ref.~\cite{Sonnentrucker2010} showed that the abundance of water vapor
within diffuse clouds in the Milky Way galaxy is remarkably constant, with
$x_{\rm H_2O} \sim 10^{-8}$, that is $\sim 0.1 \%$ of the available
oxygen. Here $x_{\rm H_2O}$ is the H$_2$O number density relative to
the total hydrogen nuclei number density.  Towards the Galactic center
this value can be enhanced by up to a factor of $\sim 3$
\cite{Monje2011, Sonnentrucker2013}.

At temperatures $\gtrsim 300$~K, H$_2$O may form directly via the
neutral-neutral reactions, O + H$_2$ $\rightarrow$ OH + H, followed by
OH + H$_2$ $\rightarrow$ H$_2$O + H.  This formation route is
particularly important in shocks, where the gas heats up to high
temperatures, and can drive most of the oxygen into H$_2$O
\cite{Draine1983, Kaufman1996}.

Temperatures of a few hundreds K are also expected in very low
metallicity gas environments, with elemental oxygen and carbon
abundances of $\lesssim 10^{-3}$ solar~\cite{2003Natur.425..812B, Omukai2005,
  Glover2014}, associated with the epochs of the first enrichment of
the ISM with heavy elements, in the first generation of galaxies at
high redshifts~\cite{LF13}.  At such low metallicities, cooling by
fine structure transitions of metal species such as the [CII]
$158$~$\mu$m line, and by rotational transitions of heavy molecules
such as CO, becomes inefficient and the gas remains warm.

Could the enhanced rate of H$_2$O formation via the neutral-neutral
sequence in such warm gas, compensate for the low oxygen abundance at
low metallicities?

Ref.~\cite{Omukai2005} studied the thermal and chemical evolution of
collapsing gas clumps at low metallicities.  They found that for
models with gas metallicities of $10^{-3}-10^{-4}$ solar, $x_{\rm
  H_2O}$ may reach $10^{-8}$, but only if the density, $n$, of the gas
approaches $10^{8}$~cm$^{-3}$.  Photodissociation of molecules by
far-ultraviolet (FUV) radiation was not included in their study.
While at solar metallicity dust-grains shield the interior of gas
clouds from the FUV radiation, at low metallicities photodissociation
by FUV becomes a major removal process for H$_2$O.  H$_2$O
photodissociation produces OH, which is then itself photodissociated
into atomic oxygen.

Ref.~\cite{Hollenbach2012} developed a theoretical model to study the
abundances of various molecules, including H$_2$O, in PDRs.  Their
model included many important physical processes, such as freezeout of
gas species, grain surface chemistry, and also photodissociation by
FUV photons.  However they focused on solar metallicity.
Intriguingly, Ref.~\cite{Bayet2009} report a water abundance close to
$10^{-8}$ in the optically thick core of their single PDR model for a
low metallicity of 10$^{-2}$ (with $n=10^3$~cm$^{-3}$).  However,
Bayet et al.~did not investigate the effects of temperature and UV
intensity variations on the water abundance in the low metallicity
regime.

Recently, a comprehensive study of molecular abundances for the bulk
ISM gas as functions of the metallicity, were studied by
Ref.~\cite{Penteado2014} and Ref.~\cite{Bialy2014};
these models however, focused on the ``low temperature" ion-molecule
formation chemistry.
  
In this section we present results for the H$_2$O abundance in low
metallicity gas environments, for varying temperatures, FUV
intensities and gas densities.  We find that for temperatures $T$ in
the range $250-350$ K, H$_2$O may be abundant, comparable to or higher
than that found in diffuse Galactic clouds, provided that the FUV
intensity to density ratio is smaller than a critical value.

\subsection{Model Ingredients}
\label{sec:method}
We calculate the abundance of gas-phase H$_2$O for low metallicity gas
parcels, that are exposed to external FUV radiation and cosmic-ray
and/or X-ray fluxes. Given our chemical network we solve the steady
state rate equations using our dedicated Newton-based solver, and
obtain $x_{\rm H_2O}$ as function of $T$ and the FUV intensity to
density ratio.

We adopt a 10$^5$~K diluted blackbody
spectrum, representative of radiation produced by massive Pop-III
stars.  The photon density in the 6-13.6 eV interval, is $n_{\gamma}
\equiv n_{\gamma, 0} I_{\rm UV}$, where $n_{\gamma, 0} = 6.5 \times
10^{-3}$~photons~cm$^{-3}$ is the photon density in the interstellar
radiation field \citep[ISRF, ][]{Draine2011}, and \iuv is the
``normalized intensity". Thus $\iuv=1$ corresponds to the FUV
intensity in the Draine ISRF.

Cosmic-ray and/or X-ray ionization drive the ion-molecule chemical
network.  We assume an ionization rate per hydrogen nucleon $\zeta$
(s$^{-1}$).  In the Galaxy, Ref.~\cite{Dalgarno2006} and
Ref.~\cite{Indriolo2012} showed that $\zeta$ lies within the
relatively narrow range $10^{-15}-10^{-16}$~s$^{-1}$. We therefore
introduce the ``normalized ionization rate" $\zeta_{-16} \equiv
(\zeta/10^{-16}$~s$^{-1})$.  The ionization rate and the FUV intensity
are likely correlated, as both should scale with the formation rate of
massive OB stars. We thus set $\zeta_{-16} = \iuv$ as our fiducial
case but also consider the cases $\zeta_{-16} = 10^{-1} \iuv$ and
$\zeta_{-16} = 10 \iuv$.

Dust shielding against the FUV radiation becomes ineffective at low
metallicities. However, self absorption in the H$_2$ lines may
significantly attenuate the destructive Lyman Werner (11.2-13.6 eV)
radiation~\cite{Draine1996, Sternberg2014} and high abundances of
H$_2$ may be maintained even at low metallicity~\cite{Bialy2014}.  In
the models presented here we assume an H$_2$ shielding column of at
least $5 \times 10^{21}$~cm$^{-2}$.  (For such conditions CO is also
shielded by the H$_2$.)  The LW flux is then attenuated by a
self-shielding factor of $f_{shield}\sim 10^{-8}$ and the H$_2$
photodissociation rate is only $5.8 \times 10^{-19} I_{\rm
  UV}$~s$^{-1}$.  With this assumption H$_2$ photodissociation by LW
photons is negligible compared to cosmic-ray and/or X-ray ionization
as long as $\iuv < 85 \zeta_{-16}$.

However, even when the Lyman Werner band is fully blocked, OH and
H$_2$O are photodissociated because their energy thresholds for
photodissociation are $6.4$ and 6 eV, respectively.  For the low
metallicities that we consider, photodissociation is generally the
dominant removal mechanism for H$_2$O and OH.  We adopt the calculated
OH and H$_2$O photodissociation rates calculated by
Ref. ~\cite{Bialy2014}.

We assume thermal and chemical steady states.  In the Milky Way, the
bulk of the ISM gas is considered to be at approximate thermal
equilibrium, set by cooling and heating processes.  We discuss the
relevant chemical and thermal time-scales in \S \ref{sub:time scales}.

Given the above mentioned assumptions, the steady state solutions for
the species abundances depend on only two parameters, the temperature
$T$ and the intensity to density ratio $\in$. Here $n_4 \equiv
(n/10^4$~cm$^{-3})$ is the total number density of hydrogen nuclei
normalized to typical molecular cloud densities.  $T$ and \in form our
basic parameter space in our study.

\subsection{Chemical Model}
\label{sub:chemical model}

We consider a chemical network of 503 two-body reactions, between 56
atomic, molecular and ionic species of H, He, C, O, S, and Si. We
assume cosmological elemental helium abundance of 0.1 relative to
hydrogen (by number).  For the metal elemental abundances we adopt the
photospheric solar abundances, multiplied by a metallicity factor $Z'$
(i.e., $Z'=1$ is solar metallicity).  In our fiducial model we assume
$Z'=10^{-3}$, but we also explore cases with $Z'=10^{-2}$ and
$Z'=10^{-4}$.  Since our focus here is on the very low metallicity
regime, where dust grains play a lesser role, we neglect any depletion
on dust grains, and dust-grain chemistry (except for H$_2$, as
discussed further below).  Direct and induced ionizations and
dissociations by the cosmic-ray / X-ray field ($\propto \zeta$) are
included.  For the gas phase reactions, we adopt the rate coefficients
given by the UMIST 2012 database \cite{McElroy2013}.

The formation of heavy molecules
relies on molecular hydrogen.  We consider two scenarios for H$_2$
formation: (a) pure gas-phase formation; and (b) gas-phase formation
plus formation on dust grains.
In gas-phase, radiative-attachment
\begin{equation}
	\label{R: H- form}
	{\rm H  \; + \;  e  \; \rightarrow \;  H^- \;  + \;  \nu}	
\end{equation}
followed by associative-detachment
\begin{equation}
	\label{R: H2 form gas}
	{\rm H^- \;  + \;  H \;  \rightarrow \;  H_2  \; + \; e} \, 
\end{equation} is the dominant H$_2$ formation route.

H$_2$ formation on the surface of dust grains is considered to be the
dominant formation channel in the Milky way.  We adopt a standard rate
coefficient~\cite{Hollenbach1971,Jura1974,Cazaux2002},
\begin{equation}
	\label{eq:R_0}
	R \ \simeq \ 3 \times 10^{-17} \ T_2^{1/2} \ Z' \ {\rm cm^3 \ s^{-1}}
\end{equation}
where $T_2 \equiv (T/100$~K$)$.  In this expression we assume that the
dust-to-gas ratio is linearly proportional to the metallicity $Z'$.
Thus, in scenario (b) H$_2$ formation on dust grains dominates even
for $Z'=10^{-3}$.  Scenario (a) is the limit where the gas-phase
channel dominates, as appropriate for dust-free environments or for
superlinear dependence of the dust-to-gas ratio on $Z'$.

\subsection{Time scales}
\label{sub:time scales}

The timescale for the system to achieve chemical steady state is
dictated by the relatively long H$_2$ formation timescale.  In the gas
phase (scenario [a]) it is
\begin{equation}
	\label{eq: time scale gas}
	t_{\rm H_2} \, = \, \frac{1}{k_2 \, n \, x_{\rm e}} \, \approx \, 8 \times 10^8 \; \zeta_{-16}^{-1/2} \ n_4^{-1/2} \, T_2^{-1} \ \rm{yr} \ ,
\end{equation} 
where $x_{\rm e}=2.4 \times 10 ^{-5} (\zeta_{-16}/n_4)^{1/2}
T_2^{0.38}$ is the electron fraction as set by ionization
recombination equilibrium, and $k_2 \approx 1.5 \times 10^{-16}
T_2^{0.67}$~cm$^3$~s$^{-1}$ is the rate coefficient for reaction
(\ref{R: H- form}).

For formation on dust grains ($\propto Z'$), the timescale is
generally shorter, with
\begin{equation}
	\label{eq: time scale grain}
	t_{\rm H_2} \, = \, \frac{1}{R \, n} \, \approx \, 10^8 \; T_2^{-1/2} \ n_4^{-1} \, \left( \frac{10^{-3}}{Z'} \right) \ \rm{yr} \ .
\end{equation}
Gas clouds with lifetimes $t \gg t_{\rm H_2}$ will reach chemical
steady state.

\begin{figure}
	\includegraphics[width=.85\textwidth]{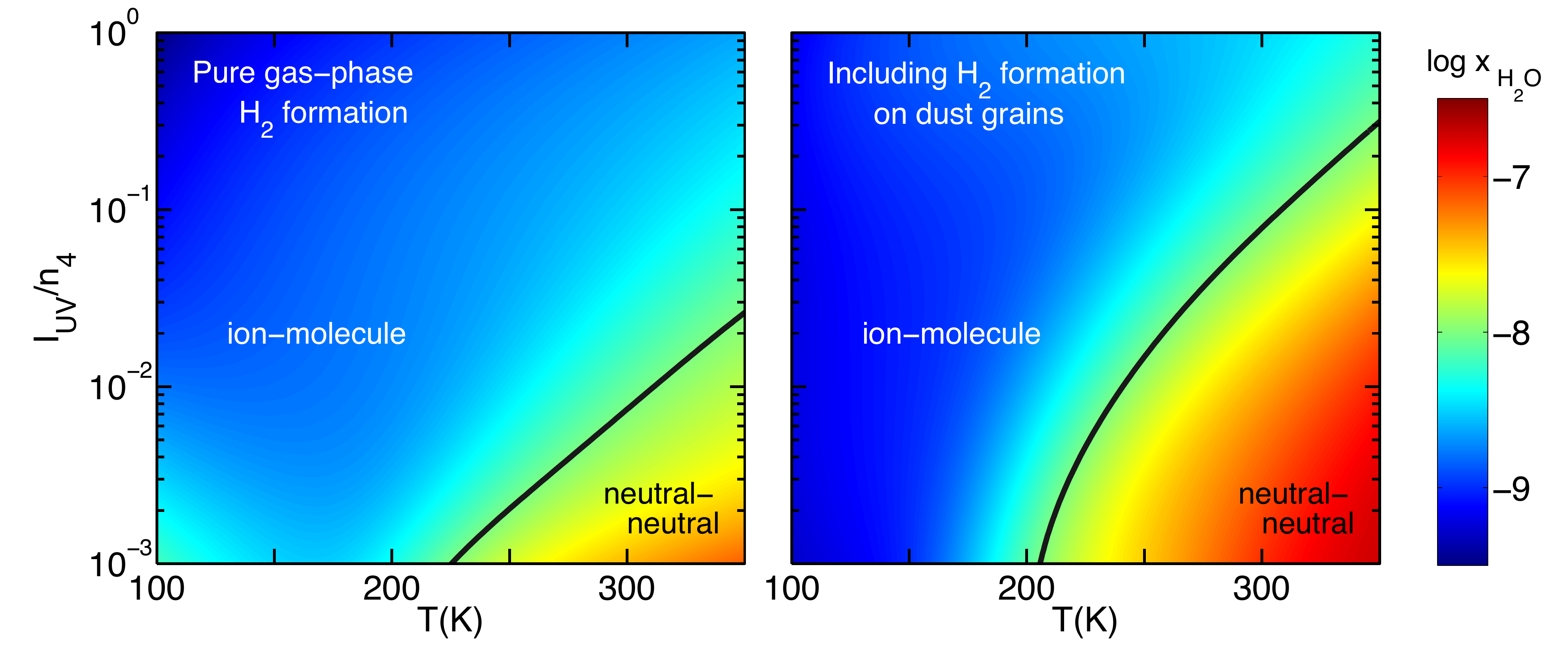}
	\caption{The fractional H$_2$O abundance $\h2o$ as a function
          of $T$ and $\in$, for $Z'=10^{-3}$ and $\zeta_{-16}/I_{\rm UV}
          =1$, assuming pure gas-phase chemistry (scenario [a] - left
          panel), and including H$_2$ formation on dust grains
          (scenario [b] - right panels).  In both panels, the solid
          line indicates the 10$^{-8}$ contour, which is a
          characteristic value for the H$_2$O gas phase abundance in
          diffuse clouds within the Milky-Way galaxy.  At high temperatures (or
          low \in values) the neutral-neutral reactions become
          effective and \h2o rises.}
	\label{fig: H2O_T_UV}
\end{figure} 

The relevant timescale for thermal equilibrium is the cooling
timescale.  For low metallicity gas with $Z'=10^{-3}$, the cooling
proceeds mainly via H$_2$ rotational excitations~\cite{Glover2014}.
If the cooling rate per H$_2$ molecule (in erg s$^{-1}$) is $W(n,T)$,
then the cooling timescale is given by
\begin{equation}
	\label{eq: time scale H2 cooling}
	t_{\rm cool} \ = \ \frac{k_B \ T}{W(n,T)} \ .
\end{equation} Here $k_B$ is the Boltzmann constant.
	For $n=10^4$~cm$^{-3}$, and $T=300$~K, $W \approx 5 \times
        10^{-25} (x_{\rm H_2}/0.1)$~erg~s$^{-1}$~\cite{LeBourlot1999},
        and the cooling time is very short $\approx 2 \times 10^3
        (0.1/x_{\rm H_2})$ yr. For densities much smaller than the
        critical density for H$_2$ cooling, $W \propto n$ and $t_{\rm
          cool} \propto 1/n$. In the opposite limit, $W$ saturates and
        $t_{\rm cool}$ becomes independent of density. We see that
        generally $t_{\rm cool} \ll t_{\rm H_2}$.
	
	Because the free fall time
	\begin{equation}
		t_{ff} \ = \ \left( \frac{3 \pi}{32 G \rho} \right)^{1/2} \ = \ 5 \times 10^5 \ n_4^{-1/2} \  {\rm yr} \ ,
	\end{equation}
	is generally much shorter than $t_{\rm H_2}$, chemical steady
        state may be achieved only in stable, non-collapsing clouds,
        with lifetimes $\gg t_{ff}$.  Obviously both $t_{\rm H_2}$ and
        $t_{\rm cool}$ must be also shorter than the Hubble time at
        the redshift of interest.

\subsection{Results}
\label{sec:Results}

\begin{figure}
	\includegraphics[width=0.85\textwidth]{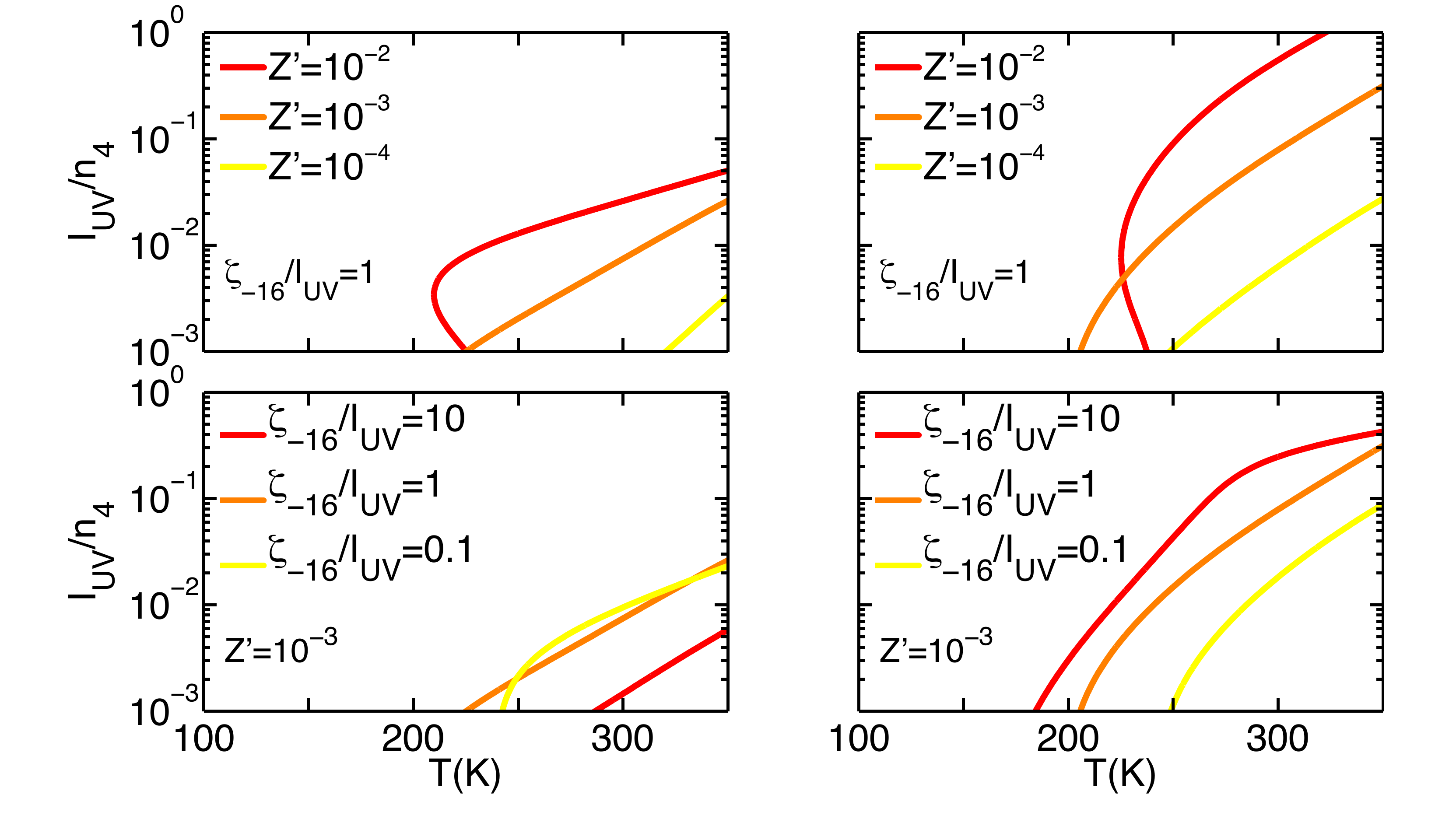}
	\caption{The $\h2o=10^{-8}$ contour, for variations in $Z'$
          (upper panels) and in $\zeta_{-16}/I_{\rm UV}$ (lower panels),
          assuming pure gas-phase chemistry (scenario [a] - left
          panels), and including H$_2$ formation on dust grains
          (scenario [b] - right panels).}
	\label{fig: variations}
\end{figure}

Next we present and discuss our results for the steady state,
gas-phase H$_2$O fraction $x_{\rm H_2O} \equiv n_{\rm H_2O}/n$, as
function of temperature $T$, and the FUV intensity to density ratio
$\in$.

\subsection{\h2o as a function of $T$ and \in}

Figure \ref{fig: H2O_T_UV} shows $\log_{10}(x_{\rm H_2O})$ contours
for the two scenarios described in \S~\ref{sub:chemical model}. In one
H$_2$ forms in pure gas-phase (scenario [a] - left panel), and in the
other H$_2$ forms also on dust-grains (scenario [b] - right
panel). Our fiducial parameters are $Z'=10^{-3}$ and
$\zeta_{-16}=\iuv$.  At the upper-left region of the parameter space,
$\h2o$ is generally low $\lesssim 10^{-9}$.  In this regime, H$_2$O
forms through the ion-molecule sequence, that is operative at low
temperatures.  In the lower right corner, the neutral-neutral
reactions become effective and \h2o rises.

In both panels, the solid line highlights the $\h2o=10^{-8}$ contour,
which resembles the H$_2$O gas phase abundance in diffuse and
translucent Milky Way clouds. This line also delineates the borderline
between the regimes where H$_2$O forms via the ``cold" ion-molecule
sequence, and the ``warm" neutral-neutral sequence.  The temperature
range at which the neutral-neutral sequence kicks-in is relatively
narrow $\sim 250-350$~K, because the neutral-neutral reactions are
limited by energy barriers that introduce an exponential dependence on
temperature.

The dependence on \in is introduced because the FUV photons
photodissociate OH and H$_2$O molecules and therefore increase the
removal rate of H$_2$O and at the same time suppress formation via the
OH + H$_2$ reaction. This gives rise to a critical value for $\in$
below which H$_2$O may become abundant.
    
For pure gas-phase H$_2$ formation (scenario [a] - left panel), the
gas remains predominantly atomic, and H$_2$O formation is less
efficient. In this case $\h2o \gtrsim 10^{-8}$ only when \in is
smaller than a critical value of
\begin{equation}
\left( I_{\rm UV}/n_4 \right)_{\rm crit}^{\rm (a)} \ = \ 2 \times 10^{-2} \ \ \ .
\end{equation}
However, when H$_2$ formation on dust is included (scenario [b] -
right panel), the hydrogen becomes fully molecular, and H$_2$O
formation is then more efficient.  In this case $\h2o$ may reach
10$^{-8}$ for $\in$ smaller than
\begin{equation}
\left( I_{\rm UV}/n_4 \right)_{\rm crit}^{\rm (b)} \ = \ 3 \times 10^{-1} \ \ \ ,
\end{equation}
an order of magnitude larger than for the pure-gas phase formation
scenario.

\subsection{Variations in $Z'$ and $\zeta_{-16}/I_{\rm UV}$}

In Figure \ref{fig: variations} we investigate the effects of
variations in the value of $Z'$ and the normalization
$\zeta_{-16}/\iuv$.  The Figure shows the $\h2o=10^{-8}$ contours for
scenarios (a) (left panels) and (b) (right panels).  As discussed
above, H$_2$O is generally more abundant in scenario (b) because the
hydrogen is fully molecular in this case, and therefore the 10$^{-8}$
contours are located at higher \in values in both right panels.

The upper panels show the effect of variations in the metallicity
value $Z'$, for our fiducial normalization $\zeta_{-16}=\iuv$.  In
both panels, the oxygen abundance rises, and $\h2o$ increases with
increasing $Z'$.  Thus at higher $Z'$, the 10$^{-8}$ contours shift to
lower $T$ and higher $\in$ and vice versa.  An exception is the
behavior of the $Z'=10^{-2}$ curve, for which the metallicity is
already high enough so that reactions with metal species dominate
H$_2$O removal for $\in \lesssim 10^{-2}$.  The increase in
metallicity then results in a {\it decrease} of the H$_2$O abundance,
and the 10$^{-8}$ contour shifts to the right.  For $\in \gtrsim
10^{-2}$ removal by FUV dominates and the behavior is similar to that
in the $Z'=10^{-3}$ and $Z'=10^{-4}$ cases.

The lower panels show the effects of variations in the ionization rate
normalization $\zeta_{-16}/\iuv$ for our fiducial metallicity value of
$Z'=10^{-3}$.  First we consider the pure gas phase formation case
(scenario [a] - lower left panel).  For the two cases
$\zeta_{-16}/\iuv = 1$ and 10$^{-1}$, the H$_2$O removal is dominated
by FUV photodissociation and therefore is independent of $\zeta$.  As
shown by Ref.~\cite{Bialy2014}, the H$_2$O formation rate is also
independent of $\zeta$ when the H$_2$ forms in the
gas-phase. Therefore \h2o is essentially independent of $\zeta$ and
the contours overlap.  For the high ionization rate $\zeta_{-16}/\iuv
= 10$, the proton abundance becomes high, and H$_2$O reactions with
H$^+$ dominate H$_2$O removal.  In this limit \h2o decreases with
$\zeta$ and the 10$^{-8}$ contour moves down.

When H$_2$ forms on dust (scenario [b] - lower right panel), the
H$_2$O formation rate via the ion-molecule sequence is proportional to
the H$^+$ and H$_3^+$ abundances, which rise with $\zeta$.  Since the
gas is molecular, the proton fraction is low and the removal is always
dominated by FUV photodissociations (independent of $\zeta$).
Therefore, in this case $\h2o$ {\it increases} with $\zeta_{-16}/\iuv$
and the curves shift up and to the left, toward lower $T$ and higher
$\in$.

\subsection{Section Summary and Implications}
\label{sec:summary}

We have demonstrated that the H$_2$O gas phase abundance may remain
high even at very low metallicities of $Z' \sim 10^{-3}$. The onset of
the efficient neutral-neutral formation sequence at $T \sim 300$~K,
may compensate for the low metallicity, and form H$_2$O in abundance
similar to that found in diffuse clouds within the Milky Way.

We have considered two scenarios for H$_2$ formation, representing two
limiting cases, one in which H$_2$ is formed in pure gas phase
(scenario [a]), and one in which H$_2$ forms both in gas-phase and on
dust grains, assuming that the dust abundance scales linearly with
$Z'$ (scenario [b]).  Recent studies by
Refs.~\cite{Galametz2011,Herrera-Camus2012,Fisher2014} suggest that
the dust abundance might decrease faster then linearly with decreasing
$Z'$.  As shown by Ref.~\cite{Bialy2014}, for $Z'=10^{-3}$ and dust
abundance that scales as $Z'^{\beta}$ with $\beta \geq 2$, H$_2$
formation is dominated by the gas phase formation channel.  Therefore
our scenario (a) is also applicable for models in which dust grains
are present, with an abundance that scales superlinearly with $Z'$.
For both scenarios (a) and (b) we have found that the neutral-neutral
formation channel yields $\h2o\gtrsim 10^{-8}$, provided that $\in$ is
smaller than a critical value.  For the first scenario we have found
that this critical value is $(I_{\rm UV}/n_4)_{\rm crit} = 2 \times
10^{2}$. For the second scenario $(I_{\rm UV}/n_4)_{\rm crit} = 3
\times 10^{-1}$.

In our analysis we have assumed that the system had reached a chemical
steady state.  For initially atomic (or ionized) gas, this assumption
offers the best conditions for the formation of molecules.  However,
chemical steady state might not always be achieved within a cloud
lifetime or even within the Hubble time.  The timescale to achieve
chemical steady state (from an initially dissociated state) is
dictated by the H$_2$ formation process, and is generally long at low
metallicities.  For $Z'=10^{-3}$ and our fiducial parameters, the
timescales for both scenarios are of order of a few 10$^{8}$ years
(see e.g. Ref.~\cite{Bell2006}), and are comparable to the age of the
Universe at redshifts of $\sim 10$.  The generically high water
abundances we find for warm conditions and low metallicities will be
maintained in dynamically evolving systems so long as they remain
H$_2$-shielded.

Our results might have interesting implications for the question of
how early could have life originated in the Universe~\cite{Loeb14}.
Our study addresses the first step of H$_2$O formation in early
enriched, molecular gas clouds.  If such a cloud is to collapse and
form a protoplanetary disk, some of the H$_2$O may make its way to the
surfaces of forming planets~\cite{vanDishoeck2014}.  However, the
question of to what extent the H$_2$O molecules that were formed in
the initial molecular clouds, are preserved all the way through the
process of planet formation, is beyond the scope of this section.

\section{An Observational Test for the Anthropic Origin of the 
Cosmological Constant}
\label{anth}

\subsection{Section Background}

The distance to Type Ia supernovae~\cite{Perlmutter,Riess} and the
statistics of the cosmic microwave background
anisotropies~\cite{Planck} provide conclusive evidence for a finite
vacuum energy density of $\rho_V= 4~{\rm keV~cm^{-3}}$ in the
present-day Universe. This value is a few times larger than the mean
cosmic density of matter today. The expected exponential expansion of
the Universe in the future (for a time-independent vacuum density)
will halt the growth of all bound systems such as galaxies and groups
of galaxies~\cite{Nagamine,Busha,Dunner}.  It will also redshift all
extragalactic sources out of detectability (except for the merger
remnant of the Milky Way and the Andromeda galaxies to which we are
bound) -- marking the end of extragalactic astronomy, as soon as the
Universe will age by another factor of ten~\cite{Loeb}.

The observed vacuum density is smaller by tens of orders of magnitude
than any plausible zero-point scale of the Standard Model of particle
physics.  Weinberg~\cite{Weinberg} first suggested that such a
situation could arise in a theory that allows the cosmological
constant to be a free parameter.  On a scale much bigger than the
observable Universe one could then find regions in which the value of
$\rho_V$ is very different. However, if one selects those regions that
give life to observers, then one would find a rather limited range of
$\rho_V$ values near its observed magnitude, since observers are most
likely to appear in galaxies as massive as the Milky-way galaxy which
assembled at the last moment before the cosmological constant started
to dominate our Universe.  Vilenkin~\cite{Vilenkin2} showed that this
so-called ``anthropic argument'' \cite{Barrow} can be used to
calculate the probability distribution of vacuum densities with
testable predictions.  This notion
\cite{Efstathiou,Martel,Tegmark,Wein,Garriga,GarrigaLivio,Tegmark1} gained
popularity when it was realized that string theory predicts the
existence of an extremely large
number~\cite{Bousso,Giddings,Maloney,Kachru}, perhaps as large as
$\sim 10^{100}$ to $10^{500}$~\cite{Ashok}, of possible vacuum
states. The resulting landscape of string vacua~\cite{Susskind} in the
``multiverse'' encompassing a volume of space far greater than our own
inflationary patch, made the anthropic argument appealing to particle
physicists and cosmologists
alike~\cite{Weinb,Polchinsky,Tegmark1}.

The time is therefore ripe to examine the prospects for an experimental
test of the anthropic argument. Any such test should be welcomed by
proponents of the anthropic argument, since it would elevate the idea to
the status of a falsifiable physical theory. At the same time, the test
should also be welcomed by opponents of anthropic reasoning, since such a
test would provide an opportunity to diminish the predictive power of the
anthropic proposal and suppress discussions about it in the scientific
literature.

{\it Is it possible to dispute the anthropic argument without visiting
  regions of space that extend far beyond the inflationary patch of
  our observable Universe?} The answer is {\it yes} if one can
demonstrate that life could have emerged in our Universe even if the
cosmological constant would have had values that are much larger than
observed. In this section we propose a set of astronomical
observations that could critically examine this issue.  We make use of
the fact that dwarf galaxies formed in our Universe at redshifts as
high as $z\sim 10$ when the mean matter density was larger by a factor
of $\sim 10^3$ than it is today\footnote{We note that although the
  cosmological constant started to dominate the mass density of our
  Universe at $z\sim 0.4$, its impact on the formation of bound
  objects became noticeable only at $z\sim 0$ or later
 ~\cite{Nagamine,Busha,Dunner}. For the purposes of our discussion, we
  therefore compare the matter density at $z\sim 10$ to that
  today. Coincidentally, the Milky-Way galaxy formed before $\rho_V$
  dominated but it could have also formed later.}~\cite{LF13}.  If
habitable planets emerged within these dwarf galaxies or their
descendents (such as old globular clusters which might be the tidally
truncated relics of early galaxies~\cite{Moore,Meylan}), then life
would have been possible in a Universe with a value of $\rho_V$ that
is a thousand times bigger than observed.

\subsection{Prior Probability Distribution of Vacuum Densities}

On the Planck scale of a quantum field theory which is unified with gravity
(such as string theory), the vacuum energy densities under discussion
represent extremely small deviations around $\rho_V=0$. Assuming that the
prior probability distribution of vacuum densities, ${\cal P}_*(\rho_V)$,
is not divergent at $\rho_V= 0$ (since $\rho_V=0$ is not favored by any
existing theory), it is natural to expand it in a Taylor series and keep
only the leading term. Thus, in our range of interest of $\rho_V$ values
\cite{Wein,Garriga,GarrigaLivio},
\begin{equation}
{\cal P_*}(\rho_V)\approx const .
\end{equation}
This implies that the probability of measuring a value equal to or smaller
than the observed value of $\rho_V$ is $\sim 10^{-3}$ if habitable planets
could have formed in a Universe with a value of $\rho_V$ that is a thousand
times bigger than observed.

Numerical simulations indicate that our Universe would cease to make new
bound systems in the near future~\cite{Nagamine,Busha,Dunner}.  A Universe
in which $\rho_V$ is a thousand times larger, would therefore make dwarf
galaxies until $z\sim 10$ when the matter density was a thousand times
larger than today.  The question of whether planets can form within these
dwarf galaxies can be examined observationally as we discuss next. It is
important to note that once a dwarf galaxy forms, it has an arbitrarily
long time to convert its gas into stars and planets, since its internal
evolution is decoupled from the global expansion of the Universe (as long
as outflows do not carry material out of its gravitational pull).

\subsection{Extragalactic Planet Searches}

Gravitational microlensing is the most effective search method for
planets beyond our galaxy.  The planet introduces a short-term
distortion to the otherwise smooth lightcurve produced by its parent
star as that star focuses the light from a background star which
happens to lie behind it \cite{Mao}--\cite{Park}. In an extensive
search for planetary microlensing signatures, a number of
collaborations named PLANET~\cite{Alb}, $\mu$FUN \cite{Yoo} and
RoboNET, are performing follow-up observations on microlensing events
which are routinely detected by the groups MOA \cite{MOA} and
OGLE~\cite{OGLE}. Many ``planetary'' events have been reported,
including a planet of a mass of $\sim 5$ Earth masses at a projected
separation of 2.6AU from a $0.2M_\odot$ M-dwarf star in the
microlensing event OGLE-2005-BLG-390Lb~\cite{Beaulieu}, and a planet
of 13 Earth masses at a projected separation of 2.3AU from its parent
star in the event OGLE-2005-BLG-169 towards the Galactic bulge -- in
which the background star was magnified by the unusually high factor
of $\sim 800$~\cite{Gould}.  Based on the statistics of these events
and the search parameters, one can infer strong conclusions about the
abundance of planets of various masses and orbital separations in the
surveyed star population~\cite{Gau,Gould,Bond}.  The technique can be
easily extended to lenses outside our galaxy and out to the Andromenda
galaxy (M31) using the method of pixel
lensing~\cite{Covone,Baltz,Chung}.  For the anthropic experiment, we
are particularly interested in applying this search technique to
lensing of background Milky-Way stars by old stars in foreground
globular clusters (which may be the tidally-truncated relics of $z\sim
10$ galaxies), or to lensing of background M31 stars by foreground
globular clusters~\cite{Huxor} or dwarf galaxies such as Andromeda
VIII~\cite{Morrison}. In addition, self-lensing events in which
foreground stars of a dwarf galaxy lens background stars of the same
galaxy, are of particular interest.  Such self-lensing events were
observed in the form of caustic-crossing binary lens events in the
Large Magellanic Cloud (LMC) and the Small Magellanic Cloud
(SMC)~\cite{rd3}. In the observed cases there is enough information to
ascertain that the most likely lens location is in the Magellanic
Clouds themselves.  Yet, each caustic-crossing event represents a much
larger number of binary lens events from the same lens population; the
majority of these may be indistinguishable from point-lens events. It
is therefore possible that some of the known single-star LMC lensing
events are due to self-lensing \cite{rd3}, as hinted by their
geometric distribution~\cite{Sahu,Gyuk}.
 
As mentioned earlier, another method for finding extra-Galactic
planets involves transit events in which the planet passes in front of
its parent star and causes a slight temporary dimming of the
star. Spectral modeling of the parent star allows to constrain both
the size and abundance statistics of the transiting
planets~\cite{Charb,Pepper}.  Existing surveys reach distance scales
of several kpc~\cite{Mall}--\cite{Urakawa} with some successful
detections \cite{OG,Konacki,Bouchy}. So far, a Hubble Space Telescope
search for transiting Jupiters in the globular cluster 47 Tucanae
resulted in no detections~\cite{Gill} [although a pulsar planet was
  discovered later by a different technique in the low-metallicity
  globular cluster Messier 4 \cite{Stein}, potentially indicating
  early planet formation].  A future space telescope (beyond the
TESS\footnote{https://tess.gsfc.nasa.gov}, Kepler
\footnote{http://kepler.nasa.gov/} and COROT
\footnote{http://smsc.cnes.fr/COROT/} missions which focus on nearby
stars) or a large-aperture ground-based facility (such as the Giant
Magellan Telescope [GMT]\footnote{http://www.gmto.org/}, the
Thirty-Meter Telescope [TMT]\footnote{http://www.tmt.org/}, or the
European Extremely Large Telescope
[EELT]\footnote{http://www.eso.org/sci/facilities/eelt}) could extend
the transit search technique to planets at yet larger distances (but
see Ref.~\cite{Pepper}). Existing searches~\cite{Charb} identified the
need for a high signal-to-noise spectroscopy as a follow-up technique
for confirming real transits out of many false events. Such follow-ups
would become more challenging at large distances, making the
microlensing technique more practical.

\subsection{Observations of Dwarf Galaxies at High-redshifts}

Our goal is to study stellar systems in the local Universe which are
the likely descendents of the early population of $z\sim 10$
galaxies~\cite{LG}. In order to refine this selection, it would be
desirable to measure the characteristic size, mass, metallicity, and
star formation histories of $z\sim 10$ galaxies (see Ref.~\cite{LF13}
for a review on their theoretically-expected properties). As already
mentioned, it is possible that the oldest globular clusters are
descendents of the first galaxies~\cite{Ricotti}.

Recently, a large number of faint early galaxies, born less than a billion
years after the big bang, have been discovered (see, e.g.~\cite{RM1,KES1}).
These include starburst galaxies with star formation rates in excess of
$\sim0.1 M_\odot~{\rm yr}^{-1}$ and dark matter halos~\cite{S1} of $\sim
10^{9-11}M_\odot$~\cite{RM1,RM2,ES1,Bow1,Bow2} at $z\sim5$--$10$.  Luminous
Ly$\alpha$ emitters are routinely identified through continuum dropout and
narrow band imaging techniques~\cite{Bow2,Bow1}. In order to study
fainter sources which were potentially responsible for reionization,
spectroscopic searches have been undertaken near the critical curves of
lensing galaxy clusters~\cite{ES1,KES1}, where gravitational
magnification enhances the flux sensitivity.  Because of the foreground
emission and opacity of the Earth's atmosphere, it is difficult to measure
spectral features other than the Ly$\alpha$ emission line from these feeble
galaxies from ground-based telescopes.

In one example, gravitational lensing by the massive galaxy cluster
A2218 allowed to detect a stellar system at $z=5.6$ with an estimated
mass of $\sim 10^6 M_\odot$ in stars~\cite{ES1}.  Detection of
additional low mass systems could potentially reveal whether globular
clusters formed at these high redshifts.  Such a detection would be
feasible with the {\it James Webb Space
  Telescope}\footnote{http://www.jwst.nasa.gov/}.  Existing designs
for future large-aperture ($>20$m) infrared telescopes (such as the
GMT, TMT, and EELT mentioned above), would also enable to measure the
spectra of galaxies at $z\sim 10$ and infer their properties.

Based the characteristics of high-$z$ galaxies, one would be able to
identify present-day systems (such as dwarf galaxies or globular clusters)
that are their likely descendents~\cite{Dolphin,Wyse} and search for
planets within them. Since the lifetime of massive stars that explode as
core-collapse supernovae is two orders of magnitude shorter than the age of
the universe at $z\sim 10$, it is possible that some of these systems would
be enriched to a high metallicity despite their old age. For example, the
cores of quasar host galaxies are known to possess super-solar metallicities
at $z\gtrsim 6$~\cite{Hamann}.

\subsection{Section Summary and Implications}

In future decades it would be technologically feasible to search for
microlensing or transit events in local dwarf galaxies or old globular
clusters and to check whether planets exist in these environments.
Complementary observations of early dwarf galaxies at redshifts $z\sim
10$ can be used to identify nearby galaxies or globular clusters that
are their likely descendents.  If planets are found in local galaxies
that resemble their counterparts at $z\sim10$, then the precise
version of the anthropic
argument~\cite{Weinberg,Vilenkin2,Efstathiou,Martel,Garriga} would be
weakened considerably, since planets could have formed in our Universe
even if the cosmological constant, $\rho_V$, was three orders of
magnitude larger.  For a flat probability distribution at these
$\rho_V$ values (which represents infinitesimal deviations from
$\rho_V=0$ relative to the Planck scale), this would imply that the
probability for us to reside in a region where $\rho_V$ obtains its
observed value is lower than $\sim 10^{-3}$. The precise version of
the anthropic argument
\cite{Weinberg,Vilenkin2,Efstathiou,Martel,Garriga} could then be
ruled-out at a confidence level of $\sim 99.9\%$, which is a
satisfactory measure for an experimental test. The envisioned
experiment resonates with two of the most active frontiers in
astrophysics, namely the search for planets and the study of
high-redshift galaxies, and if performed it would have many side
benefits to conventional astrophysics.

We note that in the hypothetical Universe with a large cosmological
constant, life need not form at $z\sim 10$ (merely 400 million years after
the big bang) but rather any time later.  Billions of years after a dwarf
galaxy had formed -- a typical astronomer within it would see the host
galaxy surrounded by a void which is dominated by the cosmological
constant.

An additional factor that enters the likelihood function of $\rho_V$
values involves the conversion efficiency of baryons into observers in
the Universe. A Universe in which observers only reside in galaxies
that were made at $z\sim 10$ might be less effective at making
observers.  The fraction of baryons that have assembled into
star-forming galaxies above the hydrogen cooling threshold by $z\sim
10$ is estimated to be $\sim10\%$~\cite{LF13}, comparable to the final
fraction of baryons that condensed into stars in the present-day
Universe~\cite{Fukugita}. It is possible that more stars formed in
smaller systems down to the Jeans mass of $\sim 10^{4-5}M_\odot$
through molecular hydrogen cooling~\cite{BL04}.  Although today most
baryons reside in a warm-hot medium of $\sim 2\times 10^6$K that
cannot condense into stars \cite{Dave,Cen}, most of the cosmic gas at
$z\sim 10$ was sufficiently cold to fragment into stars as long as it
could have cooled below the virial temperature of its host
halos~\cite{LF13}.  The star formation efficiency can be
inferred~\cite{Dolphin} from dynamical measurements of the star and
dark matter masses in local dwarfs or globulars that resemble their
counterparts at $z\sim 10$.  If only a small portion of the cosmic
baryon fraction ($\Omega_b/\Omega_m$) in dwarf galaxies is converted
into stars, then the probability of obtaining habitable planets would
be reduced accordingly. Other physical factors, such as metallicty,
may also play an important role~\cite{Livio99}.  Preliminary evidence
indicates that planet formation favors environments which are abundant
in heavy elements~\cite{Fischer}, although notable exceptions
exist~\cite{Stein}.

Unfortunately, it is not possible to infer the planet production efficiency
for an alternative Universe purely based on observations of our
Universe. In our Universe, most of the baryons which were assembled into
galaxies by $z\sim 10$ are later incorporated into bigger galaxies.  The
vast majority of the $z\sim 10$ galaxies are consumed through hierarchical
mergers to make bigger galaxies; isolated descendents of $z\sim 10$
galaxies are rare among low-redshift galaxies. At any given redshift below
10, it would be difficult to separate observationally the level of planet
formation in our Universe from the level that would have occurred otherwise
in smaller galaxies if these were not consumed by bigger galaxies within a
Universe with a large vacuum density, $\rho_V$.  In order to figure out the
planet production efficiency for a large $\rho_V$, one must adopt a
strategy that mixes observations with theory.  Suppose we observe today the
planet production efficiency in the descendents of $z\sim 10$ galaxies. One
could then use numerical simulations to calculate the abundance that these
galaxies would have had today if $\rho_V$ was $\sim 10^3$ times bigger than
its observed value. This approach takes implicitely into account the
possibility that planets may form relatively late (after $\sim$10 Gyr) within
these isolated descendents, irrespective of the value of $\rho_V$. The late
time properties of gravitationally-bound systems are expected to be
independent of the value of $\rho_V$.

In our discussion, we assumed that as long as rocky planets can form at
orbital radii that allow liquid water to exist on their surface (the
so-called {\it habitable zone}~\cite{Kasting}), life would develop over
billions of years and eventually mature in intelligence.  Without a better
understanding of the origin of intelligent life, it is difficult to assess
the physical conditions that are required for intelligence to emerge beyond
the minimal requirements stated above. If life forms early then
civilizations might have more time to evolve to advanced levels. On the
other hand, life may be disrupted more easily in early galaxies because of
their higher density (making the likelihood of stellar encounters higher)
\cite{Tegmark1,Garriga}, and so it would be useful to determine the
environmental density observationally.  In the more distant future, it
might be possible to supplement the study proposed here by the more
adventurous search for radio signals from intelligent civilizations beyond
the boundaries of our galaxy. Such a search would bring an extra
benefit. If the anthropic argument turns out to be wrong and intelligent
civilizations are common in nearby dwarf galaxies, then the older more
advanced civilizations among them might broadcast an explanation for why
the cosmological constant has its observed value.

\section{The Relative Likelihood of Life as a Function of Cosmic Time}
\label{likelihood}

\subsection{Section Background} 

Currently, we only know of life on Earth.  The Sun formed $\sim 4.6$
Gyr ago and has a lifetime comparable to the current age of the
Universe. But the lowest mass stars near the hydrogen burning
threshold at $0.08M_\odot$ could live a thousand times longer, up to
10 trillion years~\cite{laughlin1997a,Rushby}.  Given that habitable
planets may have existed in the distant past and might exist in the
distant future, it is natural to ask: {\it what is the relative
  probability for the emergence of life as a function of cosmic time?}
In this section, we answer this question conservatively by restricting
our attention to the context of ``life as we know it'' and the
standard cosmological model ($\Lambda$CDM).\footnote{We address this
  question from the perspective of an observer in a single comoving
  Hubble volume formed after the end of inflation. As such we do not
  consider issues of self-location in the multiverse, nor of the
  measure on eternally inflating regions of space-time. We note,
  however, that any observers in a post-inflationary bubble will by
  necessity of the eternal inflationary process, only be able to
  determine the age of their own bubble. We therefore restrict our
  attention to the question of the probability distribution of life in
  the history of our own inflationary bubble.} Note that since the
probability distribution is normalized to have a unit integral, it
only compares the relative importance of different epochs for the
emergence of life but does not calibrate the overall likelihood for
life in the Universe. This feature makes our results robust to
uncertainties in normalization constants associated with the
likelihood for life on habitable planets.

Next we express the relative likelihood for the appearance of life as
a function of cosmic time in terms of the star formation history of
the Universe, the stellar mass function, the lifetime of stars as a
function of their mass, and the probability of Earth-mass planets in
the habitable zone of these stars. We define this likelihood within a
fixed comoving volume which contains a fixed number of baryons.  In
predicting the future, we rely on an extrapolation of star formation
rate until the current gas reservoir of galaxies is depleted.

\subsection{Formalism}
\label{Sec:2}

\subsubsection{Master Equation}

We wish to calculate the probability $dP(t)/dt$ for life to form on
habitable planets per unit time within a fixed comoving volume of the
Universe. This probability distribution should span the time interval
between the formation time of the first stars 
%($\sim 30$~Myr after the Big Bang) 
and the maximum lifetime of all stars that were ever made
($\sim 10$~Tyr).

The probability $dP(t)/dt$ involves a convolution of the star
formation rate per comoving volume, $\dot{\rho}_*(t')$, with the
temporal window function, $g(t-t',m)$, due to the finite lifetime of
stars of different masses, $m$, and the likelihood, $\eta_{\rm Earth}(m)$,
of forming an Earth-mass rocky planet in the habitable zone (HZ) of
stars of different masses, given the mass distribution of stars,
$\xi(m)$, times the probability, $p({\rm life|HZ})$, of actually having
life on a habitable planet. With all these ingredients, the relative
probability per unit time for life within a fixed comoving volume can
be written in terms of the double integral,
\begin{equation}
	\dfrac{dP}{dt} (t) = \dfrac{1}{N} \int\limits_0^{t} dt'
        \int\limits_{m_{min}}^{m_{max}} dm' \xi(m') \dot{\rho}_*
        (t',m') \eta_{\text{Earth}}(m') p(\text{life} | \text{HZ})
        g(t-t', m'),
	\label{eq:prob}
\end{equation}
where the pre-factor $1/N$ assures that the probability
distribution is normalized to a unit integral over all times.  The
window function, $g(t-t',m)$, determines whether a habitable planet
that formed at time $t'$ is still within a habitable zone at time
$t$. This function is non-zero within the lifetime $\tau_*(m)$ of each
star, namely $g(t-t', m) = 1$ if $0<(t - t') < \tau_*(m)$, and zero
otherwise.  The quantities $m_{min}$ and $m_{max}$ represent the
minimum and maximum masses of viable host stars for habitable planets,
respectively. Below we provide more details on each of the various
components of the above master equation.

\subsubsection{Stellar Mass Range}

Life requires the existence of liquid water on the surface of
Earth-mass planets during the main stage lifetime of their host
star. These requirements place a lower bound on the lifetime of the
host star and thus an upper bound on its mass.

There are several proxies for the minimum time needed for life to
emerge. Certainly the star must live long enough for the planet to
form, a process which took $\sim 40$ Myr for
Earth~\cite{NatureLife}. Moreover, once the planet formed, sufficient
cooling must follow to allow the condensation of water on the planet's
surface. The recent discovery of the earliest crystals, Zircons,
suggest that these were formed during the Archean era, as much as 160
Myr after the planet formed~\cite{Geology}. Thus, we arrive at a
conservative minimum of 200 Myr before life could form - any star
living less than this time could not host life on an Earth-like
planet. At the other end of the scale, we find that the earliest
evidence for life on Earth comes from around 800 Myr after the
formation of the planet~\cite{NatureLife}, yielding an upper bound on
the minimum lifetime of the host star. For the relevant mass range of
massive stars, the lifetime, $\tau_*$, scales with stellar mass, $m$,
roughly as $(\tau_*/\tau_\odot) = (m/M_\odot)^{-3}$, where $\tau_\odot
\approx 10^{10}\;\text{yr}$. Thus, we find that the maximum mass of a
star capable of hosting life ($m_{max}$) is in the range $2.3$--$3.7
M_\odot$. Due to their short lifetimes and low abundances, high mass
stars do not provide a significant contribution to the probability
distribution, $dP(t)/dt$, and so the exact choice of the upper mass
cutoff in the above range is unimportant. The lowest mass stars above
the hydrogen burning threshold have a mass $m=0.08 M_\odot$.

\subsubsection{Time Range}

The 
%first stars 
%formed at a redshift of $z\sim
%70$, about $30$ Myr after the Big
%Bang~\cite{Loeb14,LF13,Fialkov,Naoz}.  supernovae from the first
stars resulted in a second generation of stars -- enriched by heavy
elements, merely a few Myr later. The theoretical expectation that the
second generation stars should have hosted planetary systems can be
tested observationally by searching for planets around metal poor
stars in the halo of the Milky Way galaxy~\cite{MashianLoeb}.

Star formation is expected to exhaust the cold gas in galaxies as soon
as the Universe will age by a factor of a few (based on the ratio
between the current reservoir of cold gas in galaxies~\cite{Fukugita}
and the current star formation rate), but low mass stars would survive
long after that. The lowest mass stars near the hydrogen burning limit
of $0.08M_\odot$, have a lifetime of order 10 trillion
years~\cite{laughlin1997a}. The probability $dP(t)/dt$ is expected to
vanish beyond that time.

\begin{figure}[!h]
	\centering
	\includegraphics[width=.5\textwidth]{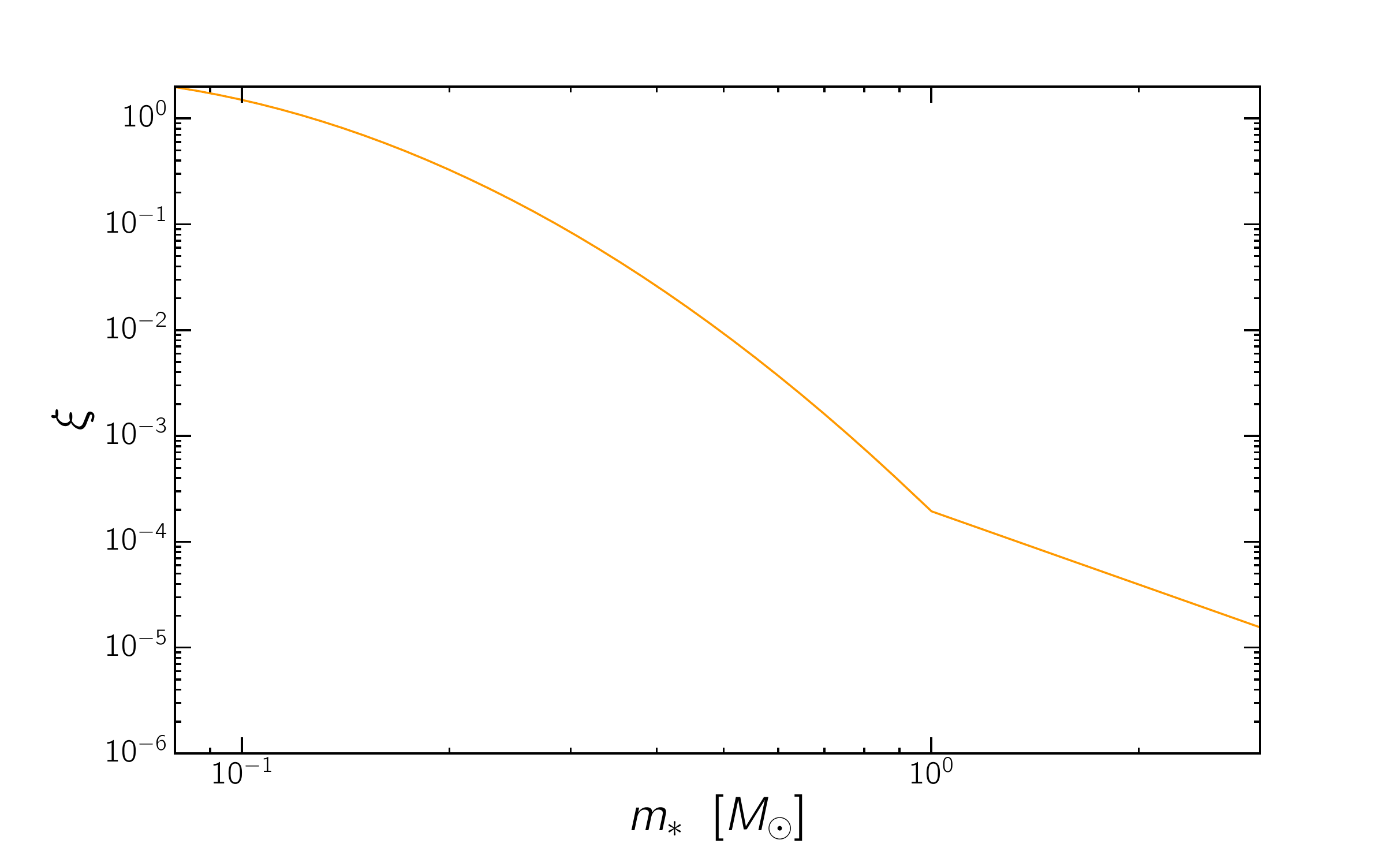}
	\caption{The Chabrier~\cite{Chabrier} mass function of stars,
          $\xi (m_*)$, plotted with a normalization integral of
          unity.}
	\label{fig:imf}
\end{figure}

\subsubsection{Initial Mass Function}

The initial mass function (IMF) of stars $\xi(m)$ is proportional to
the probability that a star in the mass range between $m$ and $m+dm$
is formed. We adopt the empirically-calibrated, Chabrier functional
form ~\cite{Chabrier}, which follows a lognormal form for masses under
a solar mass, and a power law above a solar mass, as follows:
\begin{equation}
	\xi(m) \propto 
	\begin{cases} 
		\left( \dfrac{m}{M_{\odot}} \right)^{-2.3} \quad\quad\quad\quad\quad\quad m > 1~M_\odot \\
		a \exp\left(-\dfrac{\ln(m/m_c)^2}{2\sigma^2}\right) \dfrac{M_\odot}{m} \quad m \leq  1~M_\odot 
		\end{cases},
\end{equation}
where $a = 790$, $\sigma = 0.69$, and $m_c = 0.08 M_\odot$. This IMF
is plotted as a probability distribution normalized to a unit
integral in Figure~\ref{fig:imf}.

For simplicity, we ignore the evolution of the IMF with cosmic time
and its dependence on galactic environment (e.g., galaxy type or
metallicity~\cite{Conroy}), as well as the uncertain dependence of the
likelihood for habitable planets around these stars on
metallicity~\cite{MashianLoeb}.

\subsubsection{Stellar Lifetime}

\begin{figure}[!h]
	\centering
	\includegraphics[width=.49\textwidth]{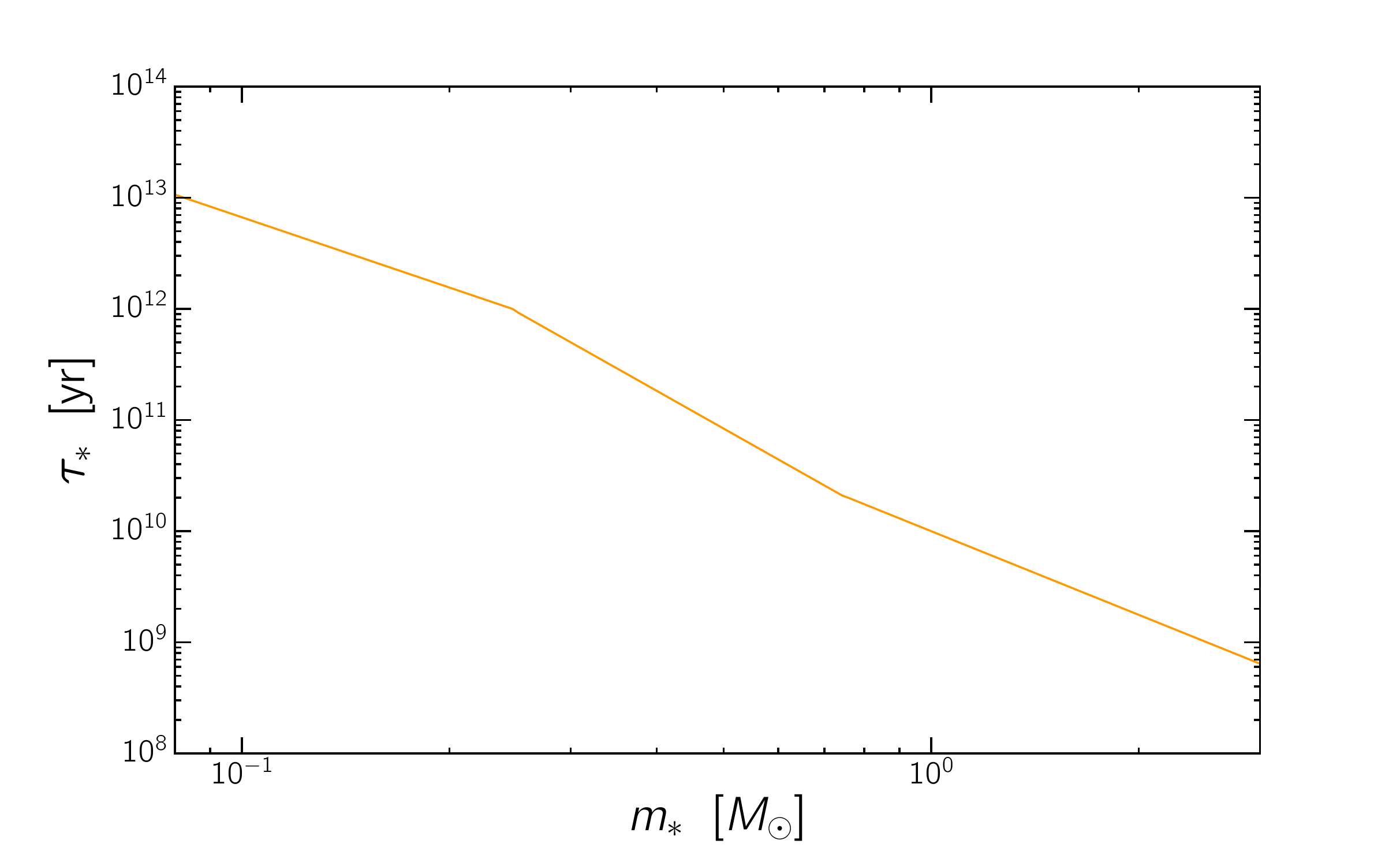}
	\caption{Stellar lifetime ($\tau_*$) as a function of 
          mass.}
	\label{fig:lt}
\end{figure}

The lifetime of stars, $\tau_*$, as a function of their mass, $m$, can
be modelled through a piecewise power-law form. For $m<0.25 M_\odot$,
we follow Ref.~\cite{laughlin1997a}. For $0.75 M_{\odot}<m<3M_\odot$,
we adopt a scaling with an average power law index of -2.5 and the
proper normalization for the Sun~\cite{Salaris}. Finally, we
interpolate in the range between 0.25 and 0.75 $M_\odot$ by fitting a
power-law form there and enforcing continuity. In summary, we adopt,
\begin{equation}
        \tau_*(m)=
	\begin{cases} 
       1.0 \times 10^{10} \; \text{yr} \left(\dfrac{m}{M_\odot}\right)^{-2.5} \quad 0.75 M_\odot < m < 3 M_\odot \\
       7.6 \times 10^{9} \; \text{yr} \left(\dfrac{m}{M_\odot}\right)^{-3.5} \quad 0.25   M_\odot < m \leq 0.75 M_\odot  \\
       5.3 \times 10^{10} \; \text{yr} \left(\dfrac{m}{M_\odot}\right)^{-2.1} \quad 0.08 M_\odot\leq m \leq  0.25 M_\odot 
	\end{cases}	.
	\label{eq:lt}
\end{equation}
This dependence is depicted in Figure~\ref{fig:lt}.

\subsubsection{Star Formation Rate}

\begin{figure}
	\centering
	\includegraphics[width=.49\textwidth]{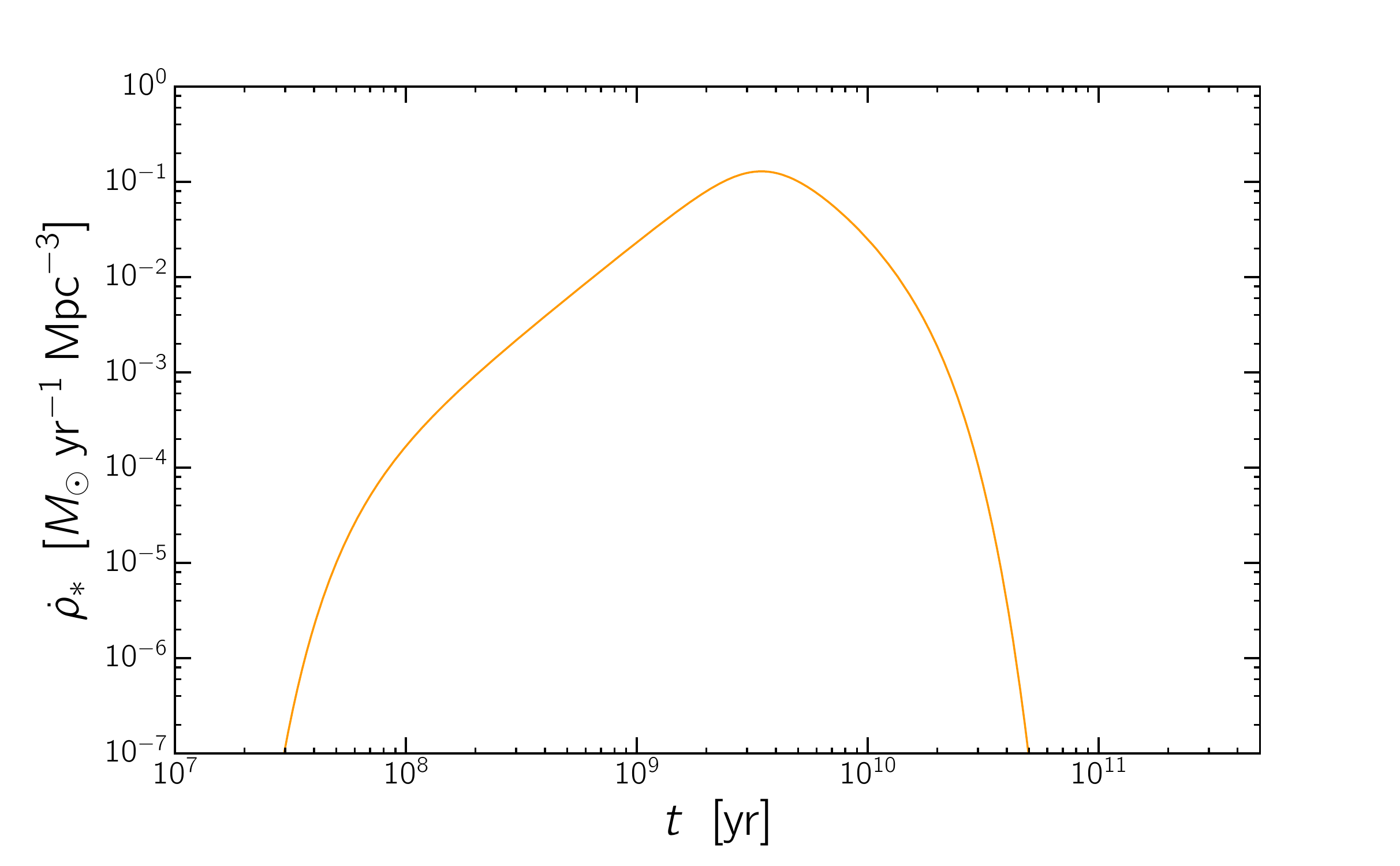}
	\includegraphics[width=.49\textwidth]{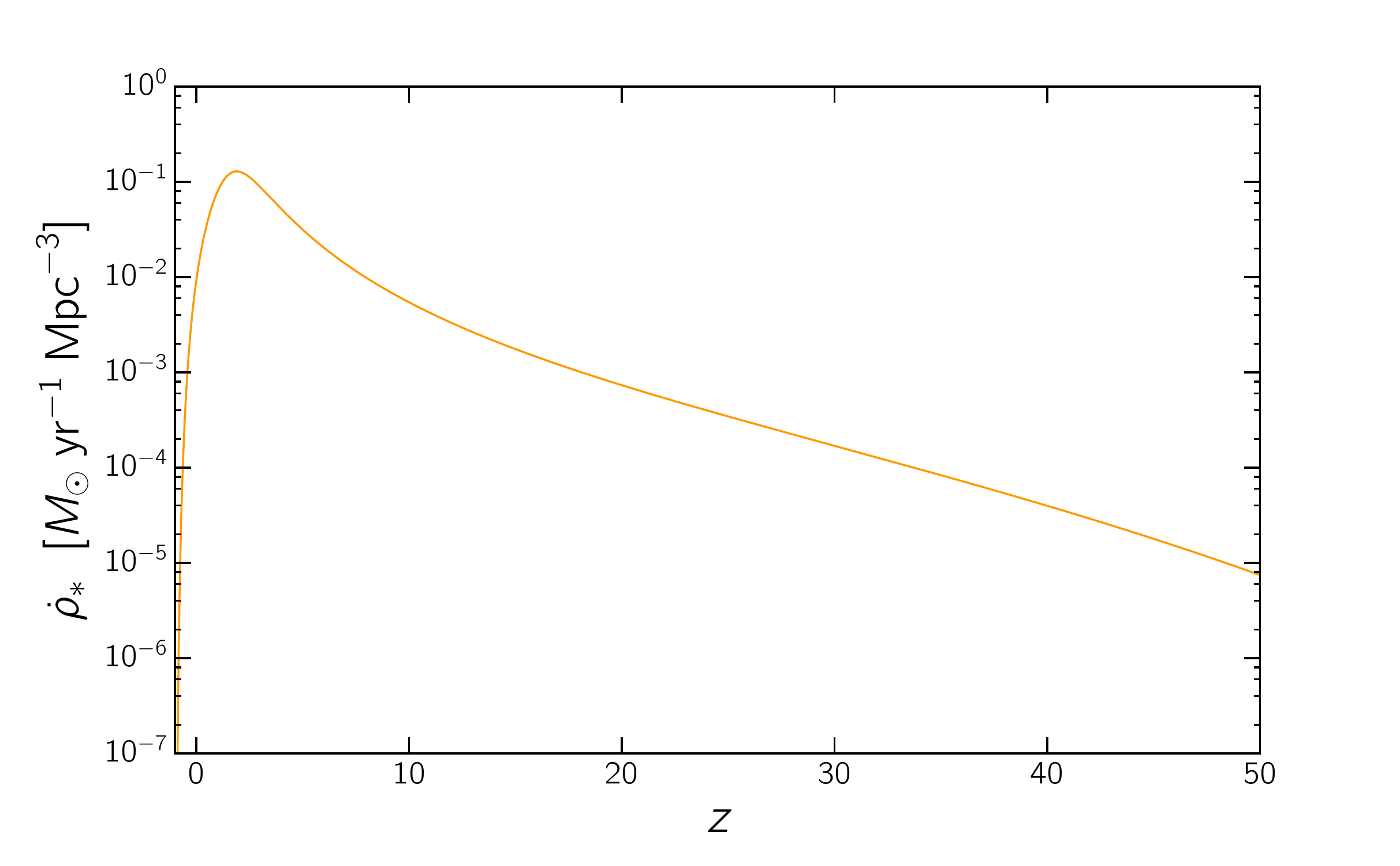}
	\caption{Star formation rate, $\dot{\rho}_*$, as a function of
          the cosmic time $t$ (left panel) and redshift $z$ (right
          panel), based on an extrapolation of a fitting function to
          existing data~\cite{MadauDickinson}.}
	\label{fig:sfr}
\end{figure}

We adopt an empirical fit to the star formation rate per comoving
volume as a function of redshift, $z$~\cite{MadauDickinson},
\begin{equation}
	\dot{\rho}_*(z) = 0.015
        \dfrac{(1+z)^{2.7}}{1+[(1+z)/2.9]^{5.6}} \; M_\odot
        \text{yr}^{-1} \text{Mpc}^{-3} ,
\end{equation}
and truncate the extrapolation to early times at the expected
formation time of the first stars~\cite{Loeb14}. We extrapolate the
cosmic star formation history to the future or equivalently negative
redshifts $-1 \le z<0$ (see, e.g. Ref.~\cite{barnes2005a}) and find
that the comoving star formation rate drops to less than $10^{-5}$ of
the current rate at 56 Gyr into the future. We cut off the star
formation at roughly the ratio between the current reservoir mass of
cold gas in galaxies~\cite{Fukugita} and the current star formation
rate per comoving volume. The resulting star formation rate as a
function of time and redshift is shown in Figure~\ref{fig:sfr}.

\subsubsection{Probability of Life on a Habitable Planet}

The probability for the existence of life around a star of a
particular mass $m$ can be expressed in terms of the product between
the probability that there is an Earth-mass planet in the star's
habitable zone (HZ) and the probability that life emerges on such a
planet: $P({\rm life}| m)=P({\rm HZ} | m)P({\rm life | HZ}) $. The
first factor, $P({\rm HZ} | m)$, is commonly labeled $\eta_{\rm
  Earth}$ in the exo-planet literature~\cite{Traub}.

\begin{figure}[!h]
	\centering 
  \includegraphics[width=0.95\textwidth]{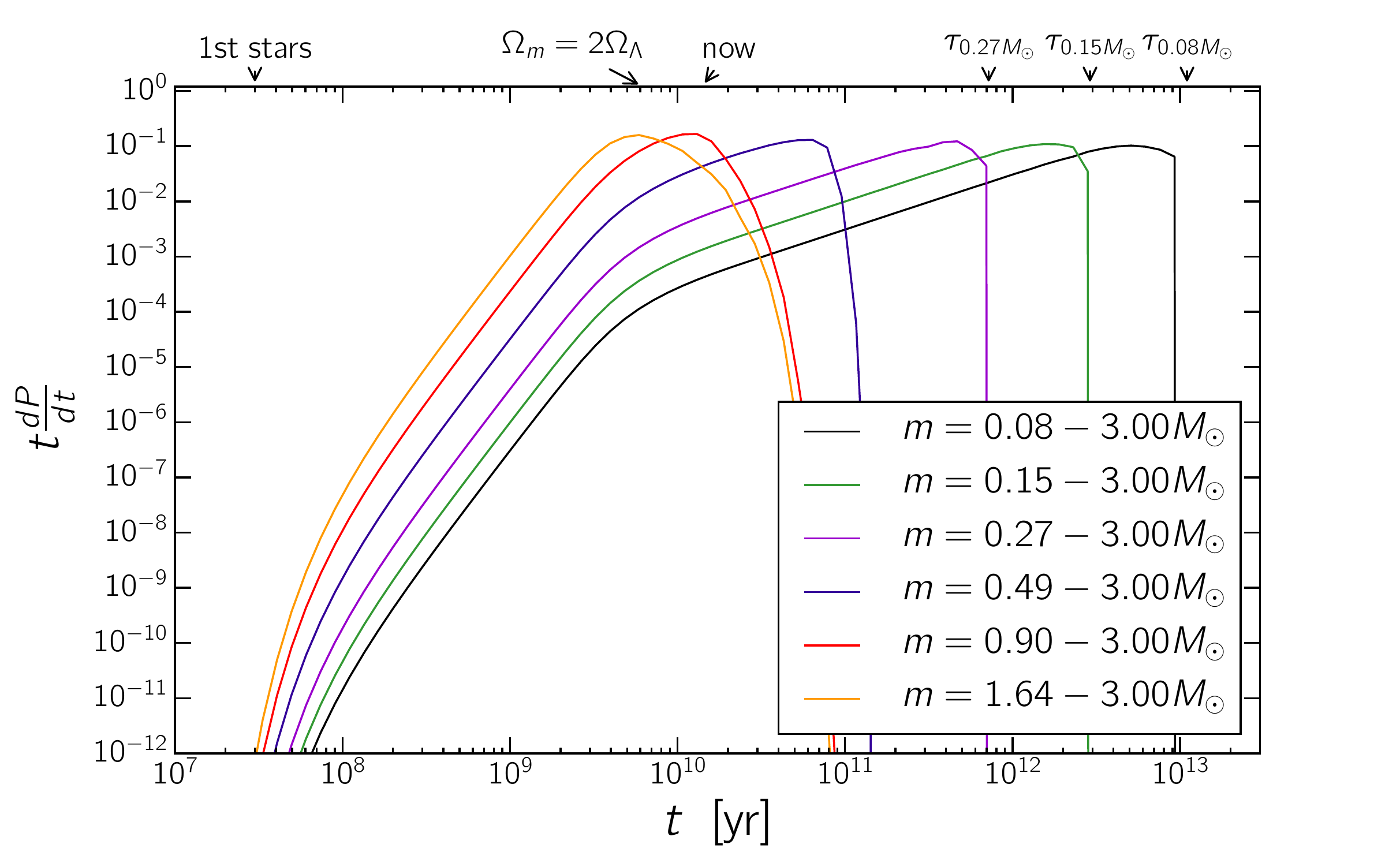} \includegraphics[width=.95\textwidth]{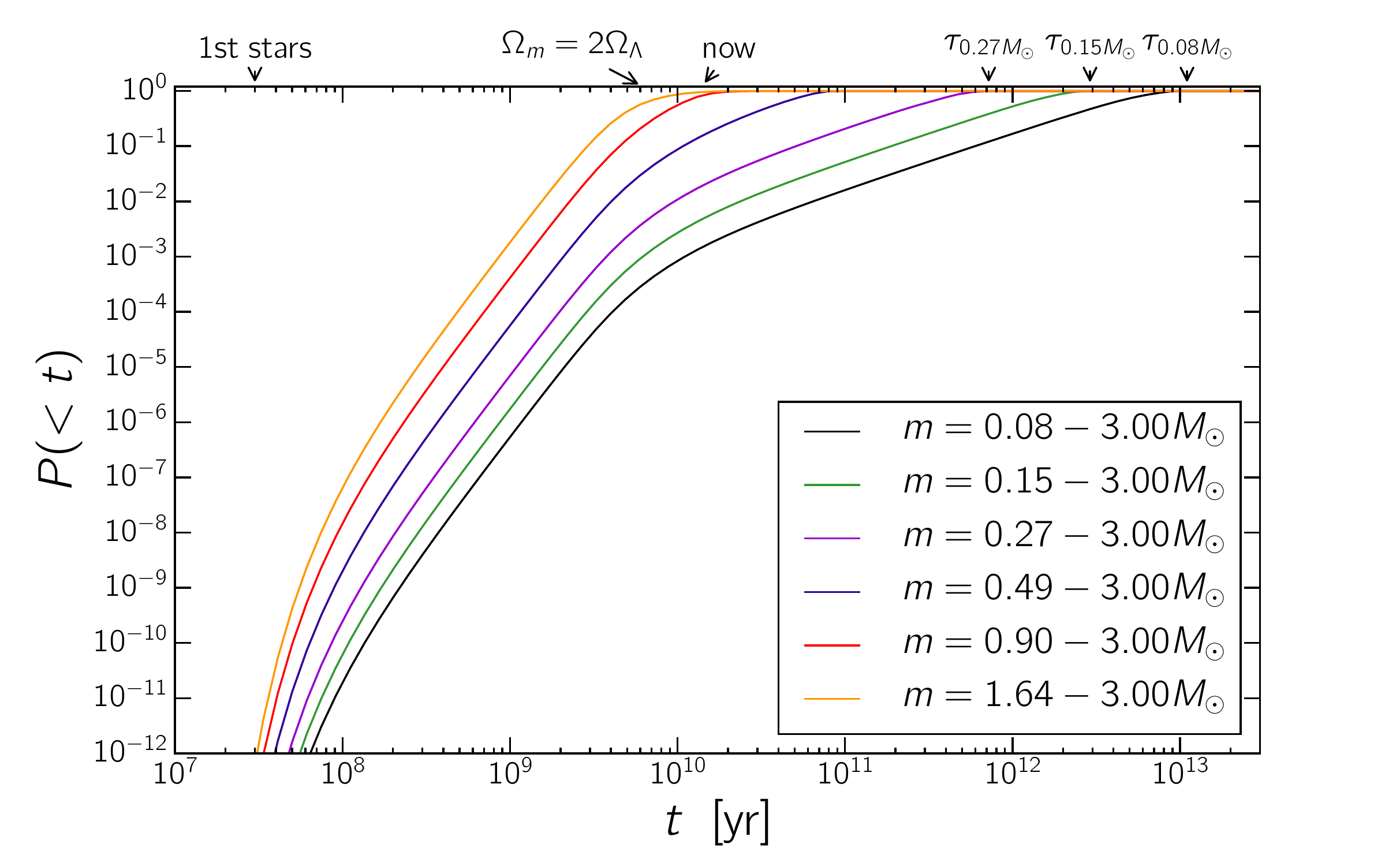} 
  \caption{Probability distribution for the emergence of life within a
  fixed comoving volme of the Universe as a function of cosmic
  time. We show the probability per log time, $tdP/dt$ (top panel) as
  well as the cumulative probability up to a time $t$, $P(<t)$ (bottom
  panel), for different choices of the minimum stellar mass, equally
  spaced in $\log m$ between $0.08M_\odot$ and $3M_\odot$. The
  contribution of stars above $3M_\odot$ to $dP(t)/dt$ is ignored due
  to their short lifetimes and low abundances. The labels on the top
  axis indicate the formation time of the first stars, the time when
  the cosmic expansion started accelerating (i.e., when the density
  parameter of matter, $\Omega_m$, was twice that of the vacuum,
  $\Omega_\Lambda$), the present time (now) and the lifetimes of stars
  with masses of $0.08M_\odot, 0.15M_\odot$ and $0.27M_\odot$.}  \label{fig:sft}
\end{figure}

Data from the NASA Kepler mission implies $\eta_{\rm Earth}$ values in
the range of $6.4^{+3.4}_{-1.1}\%$ for stars of approximately a solar
mass ~\cite{EtaEarthSilburt,Petigura,Marcy} and $24^{+18}_{-8}\%$ for
lower mass M-dwarf stars~\cite{EtaEarthDC}.  The result for solar mass
stars is less robust due to lack of identified Earth-like planets at
high stellar masses. Owing to the large measurement uncertainties, we
assume a constant $\eta_{\rm Earth}$ within the range of stellar
masses under consideration. The specific constant value of $\eta_{\rm
  Earth}$ drops out of the calculation due to the normalization factor
$N$.

There is scope for considerable refinement in the choice of the second
factor $p(\text{life}|\text{HZ})$. One could suppose that the
probability of life evolving on a planet increases with the amount of
time that the planet exists, or that increasing the surface area of
the planet should increase the likelihood of life beginning.  However,
given our ignorance we will set this probability factor to a constant,
an assumption which can be improved upon by statistical data from
future searches for biosignatures in the molecular composition of the
atmospheres of habitable
planets~\cite{Seager,LoebMaoz,Mercedes,LinLoeb}. In our simplified
treatment, this constant value has no effect on $dP(t)/dt$ since its
contribution is also cancelled by the normalization factor $N$.

\subsection{Results}
\label{Sec:3}

The top and bottom panels in Figure~\ref{fig:sft} show the probability
per log time interval $tdP(t)/dt=dP/d \ln t$ and the cumulative
probability $P(<t)=\int_0^t [dP(t')/dt'] dt'$ based on
equation~(\ref{eq:prob}), for different choices of the low mass cutoff
in the distribution of host stars for life-hosting planets (equally
spaced in $\ln m$). The upper stellar mass cutoff has a negligible
influence on $dP/d \ln t$, due to the short lifetime and low abundance
of massive stars. In general, $dP/d\ln t$ cuts off roughly at the
lifetime of the longest lived stars in each case, as indicated by the
upper axis labels. For the full range of hydrogen-burning stars,
$dP(t)/d \ln t$ peaks around the lifetime of the lowest mass stars $t\sim
10^{13}~{\rm yr}$ with a probability value that is a thousand times
larger than for the Sun, implying that life around low mass stars in
the distant future is much more likely than terrestrial life around
the Sun today.

\subsection{Section Summary and Implications}
\label{Sec:4}

Figure~\ref{fig:sft} implies that the probability for life per
logarithm interval of cosmic time, $dP(t)/d\ln t$, has a broad distribution in
$\ln t$ and is peaked in the future, as long as life is likely around
low-mass stars. High mass stars are shorter lived and less abundant
and hence make a relatively small contribution to the probability
distribution.

Future searches for molecular biosignatures (such as O$_2$ combined
with CH$_4$) in the atmospheres of planets around low mass
stars~\cite{Seager,LoebMaoz,Mercedes} could inform us whether life
will exist at late cosmic times~\cite{Koppa}. If we insist that life
near the Sun is typical and not premature, i.e. require that the peak
in $dP(t)/d\ln t$ would coincide with the lifetime of Sun-like stars
at the present time, then we must conclude that the physical
environments of low-mass stars are hazardous to life (see
e.g. Ref.~\cite{raymond2007a}).  This could result, for example, from
the enhanced UV emission and flaring activity of young low-mass stars,
which is capable of stripping rocky planets of their
atmospheres~\cite{Owen}.

Values of the cosmological constant below the observed one should not
affect the probability distribution, as they would introduce only mild
changes to the star formation history due to the modified formation
history of galaxies~\cite{Nagamine,Busha}. However, much larger values
of the cosmological constant would suppress galaxy formation and
reduce the total number of stars per comoving volume~\cite{Loeb06},
hence limiting the overall likelihood for life altogether~\cite{Weinberg}.

Our results provide a new perspective on the so-called ``coincidence
problem'', {\it why do we observe $\Omega_m \sim
  \Omega_\Lambda$?}~\cite{carter1983a} The answer comes naturally if
we consider the history of Sun-like star formation, as the number of
habitable planets peaks around present time for $m \sim 1M_\odot$. We
note that for the majority of stars, this coincidence will not exist
as $dP(t)/dt$ peaks in the future where $\Omega_m \ll
\Omega_\Lambda$. The question is then, why do we find ourselves
orbiting a star like the Sun now rather than a lower mass star in the
future?

We derived our numerical results based on a conservative set of
assumptions and guided by the latest empirical data for the various
components of equation (\ref{eq:prob}). However, the emergence of life
may be sensitive to additional factors that were not included in our
formulation, such as the existence of a moon to stabilize the climate
on an Earth-like planet~\cite{Laskar}, the existence of asteroid
belts~\cite{martin2013a}, the orbital structure of the host planetary
system (e.g. the existence of nearby giant planets or orbital
eccentricity), the effects of a binary star companion~\cite{Haghi},
the location of the planetary system within the host
galaxy~\cite{Linew}, and the detailed properties of the host galaxy
(e.g. galaxy type~\cite{Conroy} or
metallicity~\cite{Livio99,Johnson}), including the environmental
effects of quasars, $\gamma$-ray bursts~\cite{Piran} or the hot gas in
clusters of galaxies. These additional factors are highly uncertain
and complicated to model and were ignored for simplicity in our
analysis.

The probability distribution $dP(t)/d \ln t$ is of particular
importance for studies attempting to gauge the level of fine-tuning
required for the cosmological or fundamental physics parameters that
would allow life to emerge in our Universe.

\acknowledgments I thank my collaborators on the work described in
this chapter, namely Rafael Batista, Shmuel Bialy, Laura Kreidberg, ,
Gonzalo Gonzalez, James Guillochon, Henry Lin, Dani Maoz, Natalie
Mashian, David Sloan, Amiel Sternberg, Ed Turner and Matias
Zaldarriaga.

\bibliographystyle{JHEP}
\bibliography{references}

\end{document}